\documentclass{article}

\usepackage[preprint]{neurips_2026}

\usepackage{enumitem}
\usepackage[hidelinks]{hyperref}
\usepackage{url}
\usepackage{booktabs}
\usepackage{amsfonts}
\usepackage{graphicx}
\usepackage{fontawesome5}
\usepackage{sidecap}
\usepackage[small]{caption}
\usepackage{subcaption}
\usepackage{amsmath}
\allowdisplaybreaks
\usepackage{amsthm}

\usepackage{wrapfig}
\usepackage{booktabs}
\usepackage{multicol}

\usepackage{float}
\usepackage{xcolor}
\usepackage{colortbl}
\usepackage{amssymb}
\usepackage{adjustbox}
\usepackage{pifont}
\usepackage{enumitem}
\usepackage[noabbrev,capitalize]{cleveref}

\usepackage[compact]{titlesec}
\titlespacing{\section}{0pt}{1ex}{0.5ex}
\titlespacing{\subsection}{0pt}{0.5ex}{0ex}
\titlespacing{\subsubsection}{0pt}{0.5ex}{0ex} 
\usepackage{setspace}
\AtBeginDocument{%
  \addtolength\abovedisplayskip{-0.25\baselineskip}%
  \addtolength\belowdisplayskip{-0.25\baselineskip}%
  \addtolength\abovedisplayshortskip{-0.25\baselineskip}%
  \addtolength\belowdisplayshortskip{-0.25\baselineskip}%
}

\usepackage{pifont}
\usepackage{xspace}
\newcommand{\circone}{\ding{172}\xspace}
\newcommand{\circtwo}{\ding{173}\xspace}
\newcommand{\circthree}{\ding{174}\xspace}
\newcommand{\circfour}{\ding{175}\xspace}

\usepackage{tabularx, booktabs}
\newcolumntype{C}{>{\centering\arraybackslash}X}
\newcolumntype{R}{>{\raggedleft\arraybackslash}X}
\newcolumntype{S}{>{\raggedleft\arraybackslash\hsize=.5\hsize}X}

\usepackage{arydshln}
\usepackage{booktabs}
\usepackage{multirow}

\usepackage{pifont}  %
\usepackage{makecell}

\newcommand{\optparens}[1]{\if\relax\detokenize{#1}\relax\else(#1)\fi}   %

\usepackage[noabbrev,capitalize]{cleveref}
\crefname{equation}{equation}{equations}
\crefname{section}{section}{sections}
\crefname{footnote}{footnote}{footnotes}   
\crefname{line}{line}{lines}   
\crefname{assumption}{assumption}{assumptions}
\crefname{lstlisting}{listing}{listings}
\Crefname{lstlisting}{Listing}{Listings}
\crefname{appendix}{Appendix}{Appendices}
\Crefname{appendix}{Appendix}{Appendices}
\usepackage{bbm}
\usepackage{appendix}

\usepackage{xcolor}  %
\usepackage{soul}
\definecolor{aigold}{RGB}{244,210, 1} 
\definecolor{aigreen}{RGB}{245, 255, 249}

\definecolor{humanpurple}{RGB}{235, 222, 240} 

\definecolor{commentgray}{RGB}{86, 101, 115}

\definecolor{light-blue}{rgb}{0.6,0.6,1}

\definecolor{aired}{RGB}{255,180,181} 
\definecolor{codepurple}{RGB}{152,0,152}

\usepackage{listings}
\usepackage{parcolumns}
\lstdefinestyle{datalogstyle}{
	basicstyle={\codefont\small},  
	xleftmargin={6pt},
        xrightmargin={6pt},
        breakindent=0pt,
	frame=tb,
	stepnumber=1,
	firstnumber=1,
	numberfirstline=true,
	tabsize=2,
	showtabs=false,
	showspaces=false,
	showstringspaces=false,
	extendedchars=true,
	breaklines=true,
	columns=fullflexible,
	keepspaces=true,
	escapeinside={@}{@},
	firstnumber=last,
	captionpos=b,
	commentstyle=\color{black!65},
	numberstyle=\tiny\color{black!65},
	stringstyle=\color{black},
	breakatwhitespace=false, 
	keepspaces=true,                 
	numbersep=5pt,                  
	showspaces=false,                
	showstringspaces=false,
	showtabs=false,
	aboveskip={0.8\baselineskip},
	belowskip={0.2\baselineskip},
	backgroundcolor=\color{aigreen},
}
\lstdefinestyle{corebjson}{
    basicstyle={\codefont\small},
    xleftmargin={6pt},
    xrightmargin={6pt},
    frame=tb,
    tabsize=2,
    showtabs=false,
    showspaces=false,
    showstringspaces=false,
    extendedchars=true,
    breaklines=true,
    columns=fullflexible,
    keepspaces=true,
    captionpos=b,
    backgroundcolor=\color{aigreen},
    aboveskip={0.8\baselineskip},
    belowskip={0.2\baselineskip},
    morestring=[b]",
    stringstyle=\color{black},
}

\definecolor{rebuttal}{RGB}{229,255,204}
\sethlcolor{rebuttal}

\newcommand{\codefont}{\fontfamily{lmtt}\selectfont}

\usepackage[textwidth=2.7cm]{todonotes}

\newcommand{\cutforspace}[1]{}

\newcommand\myshade{85}
\colorlet{myurlcolor}{blue}
\hypersetup{
  citecolor  = black,
  urlcolor   = myurlcolor!\myshade!black,
  colorlinks = true,
}

\title{Beyond Retrieval: A Multitask Benchmark and Model for Code Search}

\author{%
  Siqiao Xue$^{\ddagger}$, Zihan Liao, Jin Qin, Ziyin Zhang, \\
  \textbf{Yixiang Mu$^{\ddagger}$, Fan Zhou$^{\ddagger}$, Hang Yu$^\dagger$}\\[0.6em]
  Ant Group \\[0.6em]
  Hangzhou, China \\[0.6em]
  \textbf{\href{https://hq-bench.github.io/coreb-page/}
     {\textcolor{blue!60!black}{\faGlobe\enspace{Website}}}
    \quad
  \href{https://github.com/hq-bench/coreb}
     {\textcolor{blue!60!black}{\faGithub\enspace{Code}}}
     \quad
    \href{https://huggingface.co/datasets/hq-bench/coreb}%
      {\textcolor{blue!60!black}{\faDatabase\enspace Data}}%
  }%
}

\begin{document}

\maketitle
{
  \renewcommand{\thefootnote}{\fnsymbol{footnote}}
  \footnotetext[1]{$^\ddagger$Work done at Alipay.\quad $^\dagger$Corresponding author.}
}

\begin{abstract}
Code search now underpins not only developer-facing tools but also modern AI coding agents (e.g., SWE-agent, OpenHands, Cursor). Yet existing benchmarks evaluate only the embedding stage, ignoring the reranker and developer-style queries that production pipelines actually use, and additionally suffer from data contamination, label noise, and degenerate binary relevance.  
In this paper, we introduce \textsc{CoREB}, a contamination-limited, multitask \underline{co}de \underline{r}etrieval and r\underline{e}ranking \underline{b}enchmark, together with a fine-tuned code reranker, that goes beyond retrieval to cover the full code search pipeline.
\textsc{CoREB} is built from counterfactually rewritten LiveCodeBench problems in five programming languages and delivered as timed releases with graded relevance judgments.
We benchmark eleven embedding models and five rerankers across three tasks: text-to-code, code-to-text, and code-to-code.
Our experiments reveal that:
\circone code-specialized embeddings dominate code-to-code retrieval (${\sim}2{\times}$ over general encoders), yet no single model wins all three tasks;
\circtwo short keyword queries, the format closest to real developer search, collapse every model to near-zero nDCG@10;
\circthree off-the-shelf rerankers are task-asymmetric, with a 12-point swing on code-to-code and no baseline net-positive across all tasks;
\circfour our fine-tuned \textsc{CoREB-Reranker} is the only reranker we evaluate that achieves consistent gains across all three tasks.
The data and model are released via the project site: \small \url{https://hq-bench.github.io/coreb-page/}.
\end{abstract}


\section{Introduction}
\label{section:intro}

Large language models (LLMs) for code~\citep{zhang2024unifying} have rapidly become foundational components of modern software engineering. Among their most impactful applications is code search, which enables developers to efficiently locate relevant code snippets, explanations, and semantically similar implementations, as exemplified by systems like GitHub Code Search~\citep{githubcodesearch}. Modern code search is rarely a single-stage process: it pairs an embedding model that maps programs into continuous vector spaces for fast first-stage retrieval with a downstream reranker that refines the final ranking. The same retrieve-and-rerank pipeline now also powers a new generation of AI coding agents---such as SWE-agent~\citep{yang2024sweagent}, OpenHands~\citep{wang2025openhands}, and IDE assistants like Cursor~\citep{cursor2024}---which rely on code search to ground their LLMs in the right files and snippets before editing or repairing code. Improvements at either stage therefore propagate beyond search itself and lift the entire AI coding pipeline.

Despite rapid progress, evaluating the full code search pipeline remains challenging. We identify four issues that limit current practice. \textbf{(D1)}\label{item:d1}~\emph{Missing reranking support.}
No existing code retrieval benchmark evaluates or provides a reranking stage. Practitioners must assemble their own pipelines using generic encoders not trained on code, which may hurt retrieval quality on code-specific tasks.

The remaining issues concern the retrieval stage specifically. Through a dataset-level analysis of CoIR~\citep{li2024coir}, the most prominent benchmark for code retrieval, we identify:
\textbf{(D2)}\label{item:d2}~\emph{Contamination and benchmark overfitting.}
CodeSearchNet~\citep{husain2020codesearchnet} and its derivatives together account for over $85\%$ of CoIR's corpus volume and have served as pre-training data for a line of code representation models (CodeBERT~\citep{feng-etal-2020-codebert}, CodeT5~\citep{wang2021codet5}, and CodeRetriever~\citep{li2022coderetriever}), many of which also serve as backbones for downstream code embedding models.
This training--evaluation overlap can inflate metrics by up to 100\%~\citep{allamanis2019adverse}; inter-dataset leakage further distorts results~\citep{hernandezlopez2024interdataset}; and \citet{karaman2025code2doc} report 15--25\% test/train near-duplication in CodeSearchNet-style data, while \citet{siddiq2024faultinstars} find evidence of data contamination in APPS.
\textbf{(D3)}\label{item:d3}~\emph{Label noise and trivial matching.}
\citet{gong2026cosqa} estimate that around 51\% of pairs in CoSQA~\citep{huang2021cosqa}, CoIR's only human-annotated dataset, are mismatched under functional verification; our own manual inspection of 80 test pairs finds approximately 60\% to be problematic.
CodeSearchNet pairs code with its own verbatim docstring, a task the original authors flag as ``overly simplistic'' due to shared authorship vocabulary~\citep{husain2020codesearchnet}, and \citet{li2024procqa} confirm produces ``documentation strings or comments rather than natural language questions,'' limiting real-world applicability.
Beyond label quality, three of ten CoIR datasets do not test code retrieval at all: CodeSearchNet-CCR randomly splits functions mid-token as a string-completion proxy, CodeFeedback-MT is dialogue continuation where prior assistant turns already contain complete solutions, and StackOverflow-QA contains virtually no code.
\textbf{(D4)}\label{item:d4}~\emph{Degenerate relevance structure.}
All ten datasets assign exactly one relevant document per query with binary scores and no explicit hard negatives, collapsing nDCG@$k$ and MRR into redundant hit-or-miss signals and making Recall@$k$ binary, a limitation the CoIR authors themselves acknowledge~\citep{li2024coir}.
Moreover, meaningful tasks cover only Python and SQL (text-to-code) or Python and C++ (code-to-code), leaving languages such as Go and Ruby entirely untested~\citep{diera2023gencodesearchnet}.
See \cref{app:coir_qrels,app:coir_cosqa,app:coir_csn,app:coir_ccr,app:coir_cfmt,app:coir_other} for a dataset-by-dataset analysis.

To fill this gap, we introduce \textsc{CoREB}, a multitask, contamination-limited \underline{co}de \underline{r}etrieval and r\underline{e}ranking \underline{b}enchmark, together with a fine-tuned code reranker, that addresses the above flaws by design and covers both stages of the code search pipeline.

\paragraph{How \textsc{CoREB} addresses D1--D4.}
To resolve \hyperref[item:d1]{D1}, we train and release a code reranker on hard negatives mined from \textsc{CoREB}'s own retrieval runs, providing a ready-to-use two-stage pipeline that improves over generic reranking on every retrieval direction we evaluate.

For \hyperref[item:d2]{D2}, rather than mining from public repositories, we build \textsc{CoREB} around counterfactually rewritten LiveCodeBench~(LCB)~\citep{jain2024livecodebench} problems and regenerated artifacts: each problem statement is rewritten to remove memorized surface forms while preserving its functional semantics, and frontier code models are re-evaluated on the refreshed instances. 

Our annotation study shows that Pass@1 consistently decreases after rewriting whenever the underlying problems fall within a model's training cutoff, indicating that some original LCB problems are partially memorized by contemporary models such as Gemini 3 Flash and Claude Sonnet 4.5.
Using the refreshed problems and their resulting test cases, we automatically construct text-to-code, code-to-text, and code-to-code tasks across five programming languages, and re-run this pipeline whenever LCB releases new problems; we elaborate on the construction pipeline in \cref{sec:construction}.



We resolve \hyperref[item:d3]{D3} by deriving relevance labels programmatically from execution outcomes: every code candidate is run against test oracles and pass/fail status determines relevance, eliminating the human annotation noise that causes ${\sim}$51\% mislabeling in CosQA~\citep{gong2026cosqa}.

Finally, we address \hyperref[item:d4]{D4} with \emph{graded} relevance judgments that mix multiple positives and explicit hard negatives in a single relevance scheme: each judgment assigns \texttt{relevance=2} to true positives, \texttt{relevance=1} to same-problem hard negatives (e.g., failed code solutions for text-to-code and code-to-code; LLM-generated noise descriptions for code-to-text), and treats unjudged items as easy negatives (\cref{tab:graded_qrel_counts}).
Across our two releases, 68\% of text-to-code queries have two or more correct solutions and code-to-code queries average 2.2 valid cross-language translations, so ranking metrics such as nDCG and Recall reflect meaningful ordering rather than a one-hit test; evaluation uses \texttt{relevance\_level=2}, so a model that retrieves a hard negative above a true positive is penalized.

\paragraph{Empirical findings.}
We evaluate eleven embedding models and five rerankers across all three tasks. Key findings include:
\circone No single model wins every task; leaderboard rankings shift across retrieval directions.
\circtwo Code-specialised training matters more than scale: a 0.5\,B code-trained model outperforms general-purpose 8\,B encoders, with the premium concentrated on code-to-code (${\sim}2{\times}$).
\circthree Two systemic failures span every model: short keyword queries collapse to near-zero nDCG@10, and low-resource languages lag by up to 0.33 points.
\circfour Reranking is high-stakes: a 12-point swing separates the best and worst baselines on code-to-code, and no off-the-shelf reranker is net-positive across all tasks. Our fine-tuned \textsc{CoREB-Reranker} is the first to achieve consistent gains on all three tasks, establishing a complete two-stage pipeline released alongside the benchmark.

\section{Related Work}
\label{sec:related_work}

Code LLMs have progressed from specialized systems like Codex~\citep{chen2021eval} to frontier models such as GPT-4.1~\citep{openai2025gpt41}, Claude 3.5~\citep{anthropic2024claude35sonnet}, and Gemini~3~Pro~\citep{gemini3pro2025}, as well as open families including StarCoder2~\citep{lozhkov2024starcoder2stackv2} and Qwen2.5-Coder~\citep{hui2024qwen25codertechnicalreport}, where strong coding capability is integrated alongside general reasoning.
For evaluation, BEIR~\citep{thakur2021beir} and MTEB~\citep{muennighoff2022mteb} cover text retrieval broadly but include few code tasks; code-specific suites like CoIR~\citep{li2024coir} and CPRet~\citep{deng2025cpret} unify multiple datasets yet inherit structural problems from their constituents: degenerate 1-to-1 qrels, label noise, contamination risk, and tasks that reduce to string matching rather than genuine retrieval (\cref{app:coir_qrels,app:coir_cosqa,app:coir_csn,app:coir_ccr,app:coir_cfmt,app:coir_other}).
On the retrieval-model side, dense encoders from CodeBERT~\citep{feng-etal-2020-codebert} to CodeSage~\citep{zhang2024code} and Qodo-Embed~\citep{qodoembed2025} achieve strong results on these benchmarks, but their behavior on broader, contamination-limited scenarios remains underexplored.
See \cref{app:related_work} for a fuller discussion.

\section{Benchmark Construction}
\label{sec:construction}

\begin{figure*}[tb]
  \centering
  \begin{subfigure}[b]{0.32\linewidth}
    \centering
    \includegraphics[width=\linewidth]{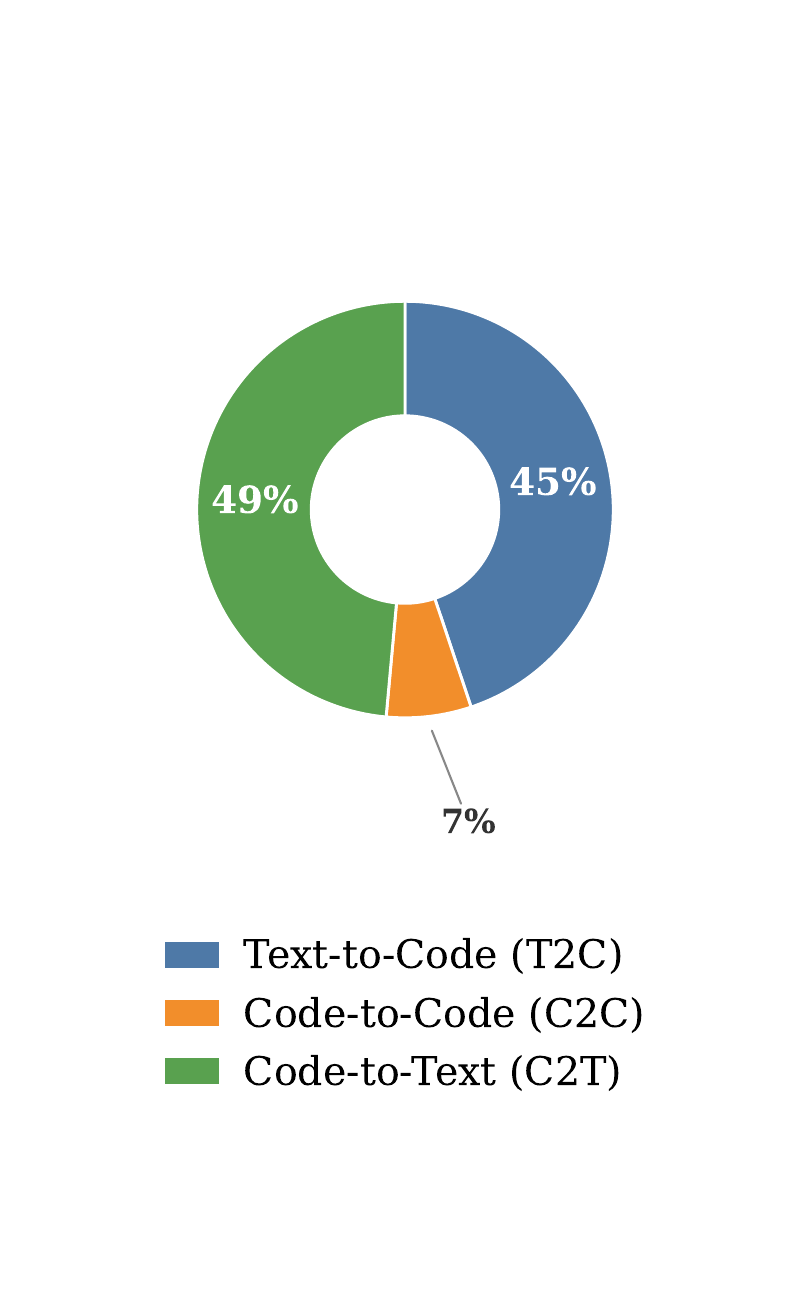}
    \caption{By main task}
    \label{fig:task_overview_a}
  \end{subfigure}
  \hfill
  \begin{subfigure}[b]{0.32\linewidth}
    \centering
    \includegraphics[width=\linewidth]{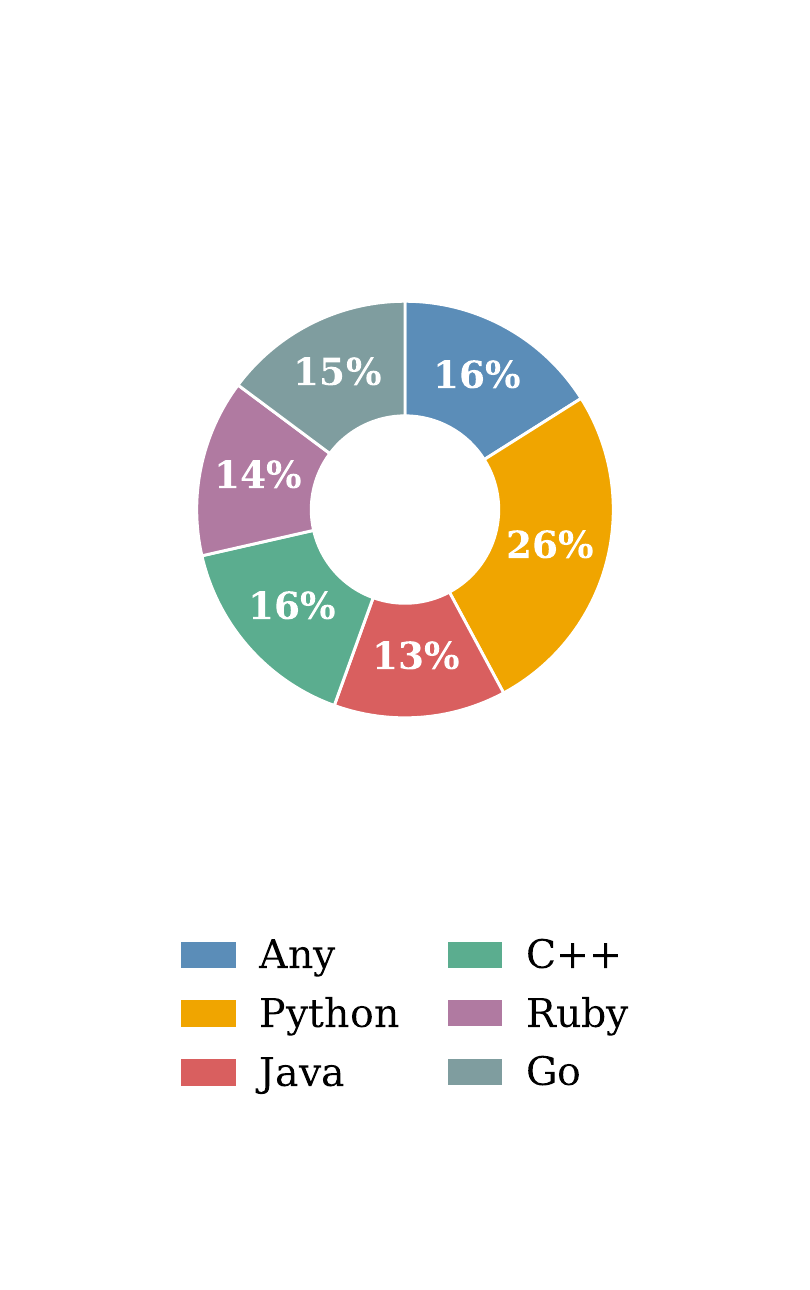}
    \caption{By language constraint}
    \label{fig:task_overview_b}
  \end{subfigure}
  \hfill
  \begin{subfigure}[b]{0.32\linewidth}
    \centering
    \includegraphics[width=\linewidth]{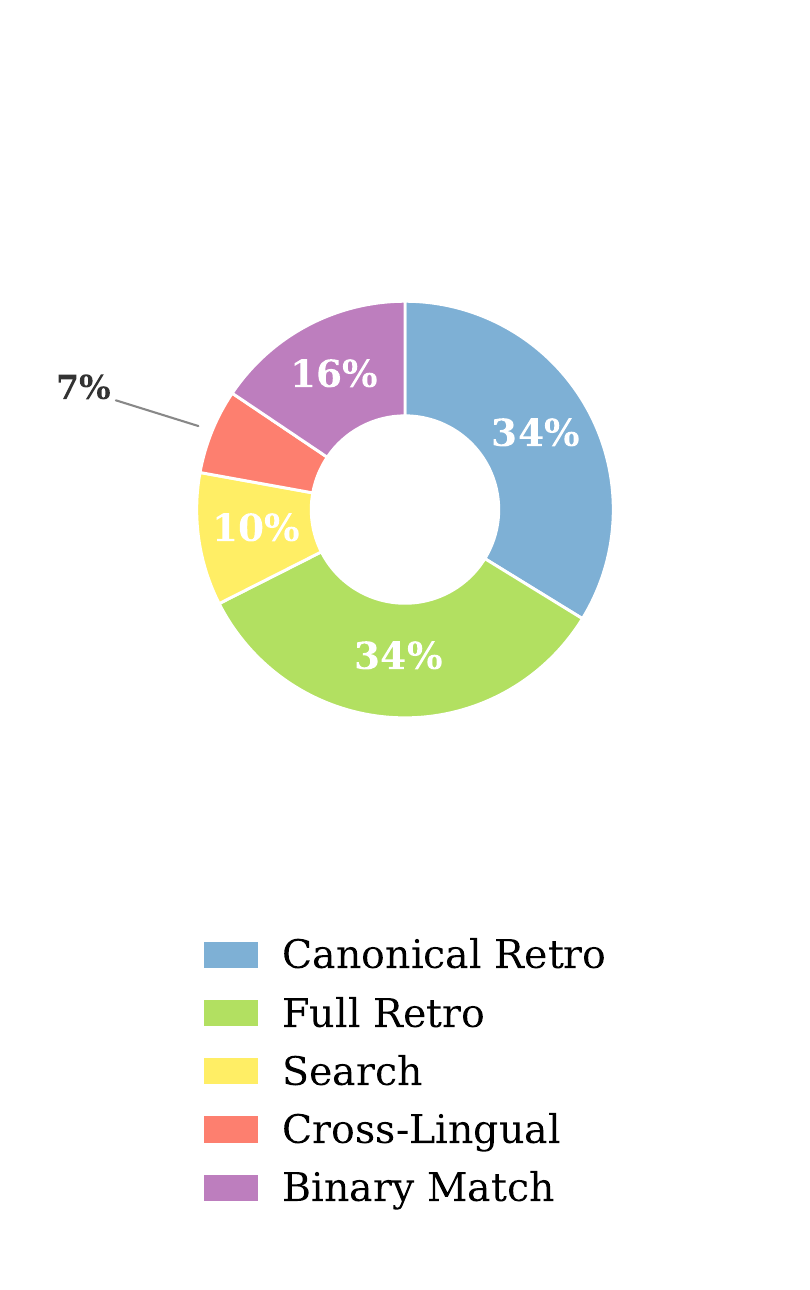}
    \caption{By subtask variant}
    \label{fig:task_overview_c}
  \end{subfigure}
  \caption{Aggregated query distribution across both \textsc{CoREB} releases
  (5{,}087 queries total). Percentages ${<}$8\% are shown outside with leader lines.
  Per-release breakdowns are in \cref{fig:task_overview_per_release}.}
  \label{fig:task_overview}
\end{figure*}

\begin{table*}[tb]
  \centering
  \label{tab:query_subtasks}
  \small
  \renewcommand{\arraystretch}{1.2}
  \setlength{\tabcolsep}{5pt}
  \begin{sc}
  \begin{tabularx}{\textwidth}{l l X r}
    \toprule
    \textbf{Main Task} & \textbf{Subtask} & \textbf{Query Description} & \textbf{Avg.\ Tokens} \\
    \midrule
    \multirow{3}{*}{Text-to-Code}
      & Canonical Retro\textsuperscript{$\dagger$}
      & Abbreviated problem descriptions & 120 \\
    & Full Retro\textsuperscript{$\dagger$}
      & Full problem statements & 431 \\
    & Search\textsuperscript{$\dagger$}\textsuperscript{$\ddagger$}
      & LLM-generated search queries & 19 \\
    \midrule
    Code-to-Code
    & Cross-Lingual
      & Code snippet (anchor) & 207 \\
    \midrule
    \multirow{3}{*}{Code-to-Text}
      & Canonical Retro\textsuperscript{$\dagger$}
      & Code solution  & 246 \\
    & Full Retro\textsuperscript{$\dagger$}
      & Code solution & 246 \\
    & Pair Match
      & Code snippet; 1 relevant doc & 252 \\
    \bottomrule
  \end{tabularx}
  \end{sc}
  \vspace{2pt}
  \caption{Query subtasks in \textsc{CoREB} with per-subtask average query length.
  Token counts use \texttt{cl100k\_base}; exact per-language counts in
  \cref{tab:query_dist_summary}; corpus statistics in \cref{tab:corpus_stats}. \footnotesize\raggedright
  \textsuperscript{$\dagger$}Each subtask has 6 language variants
  (Any + Python, Java, C++, Ruby, Go); bar colours in \cref{fig:task_overview}
  show this breakdown.
  \textsuperscript{$\ddagger$}The \emph{Search} subtask refers to a specific query type (short developer-style keyword queries); it is distinct from the broader ``code search'' used in the paper title, which encompasses the full retrieve-then-rerank pipeline.}
\end{table*}

\vspace{-0.2cm}

\begin{figure}[t]
    \centering
    \includegraphics[width=\linewidth]{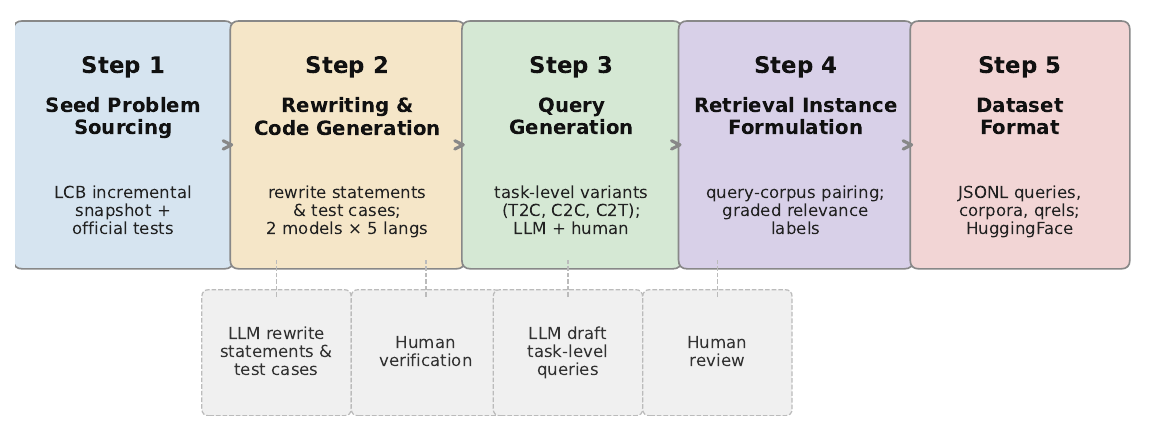}
    \caption{Benchmark construction pipeline. Each step is detailed in the corresponding paragraph below.}
    \label{fig:pipeline}
\end{figure}

Our benchmark is constructed through a multi-stage pipeline (\cref{fig:pipeline}) that converts recent competitive-programming problems into a suite of code representation and retrieval tasks. The design emphasizes contamination resistance, semantic diversity, and evaluation realism. 


\paragraph{Step 1: Seed problem sourcing from LCB.}
We source problems from LCB, a continuously updated benchmark that mitigates contamination via temporal filtering.
Mirroring LCB's incremental model, \textsc{CoREB} is published as timed releases (\cref{tab:coreb_releases}):\footnote{\small Source test splits: \url{https://huggingface.co/datasets/livecodebench/code_generation_lite/blob/main/test5.jsonl}, \url{https://huggingface.co/datasets/livecodebench/code_generation_lite/blob/main/test6.jsonl}} \texttt{v202602} (covering contest problems from Sep~2024 to Jan~2025) and \texttt{v202603} (Jan to Apr~2025).
Dataset statistics are aggregated across both releases unless a per-release breakdown is shown.

\begin{table}[tb]
\centering
\small
\setlength{\tabcolsep}{5pt}
\begin{sc}
\begin{tabular}{l ccc rr}
\toprule
& \multicolumn{3}{c}{LiveCodeBench} & \multicolumn{2}{c}{Synthetic} \\
\cmidrule(lr){2-4}\cmidrule(lr){5-6}
Release & Source & Contest period & \makecell{\# Seed\\Problems} & \makecell{\# Code\\Corpus} & \makecell{\# Text\\Corpus} \\
\midrule
v202602 & release\_v5 & Sep'24--Jan'25 & 167 & 1{,}670 & 835 \\
v202603 & release\_v6 & Jan'25--Apr'25 & 175 & 1{,}744 & 875 \\
\midrule
Aggregated & & & 342 & 3{,}414 & 1{,}710 \\
\bottomrule
\end{tabular}
\end{sc}
\vspace{2pt}
\caption{\textsc{CoREB} releases. \emph{Contest period} gives the date range of LiveCodeBench problems included in that snapshot. \emph{\# Code Corpus} counts generated solutions across five languages (2 models $\times$ 5 languages $\times$ \# seed problems); \emph{\# Text Corpus} counts original descriptions plus LLM-generated hard negatives (4 per problem). The two releases draw from disjoint LiveCodeBench snapshots, so the Aggregated row is a straight sum of distinct records. Per-task query counts are in \cref{tab:query_dist_summary}.}
\label{tab:coreb_releases}
\end{table}

\begin{wrapfigure}{r}{0.42\textwidth}
    \centering
    \vspace{-12pt}
    \includegraphics[width=0.42\textwidth]{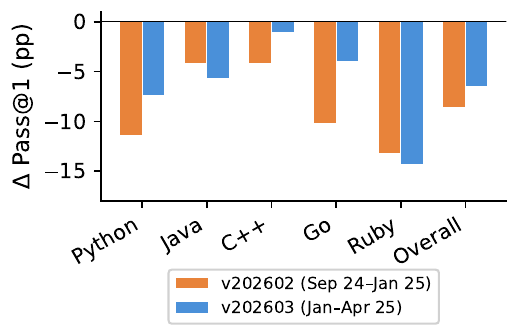}
    \caption{Pass@1 change (pp) after rewriting for Gemini~3~Flash across two releases covering different contest windows.}
    \label{fig:annotation_pass1}
    \vspace{-8pt}
\end{wrapfigure}

\paragraph{Step 2: Counterfactual rewriting and code generation.}
We apply light counterfactual rewriting~\citep{wu2024cofca} to each problem's statement \emph{and} test cases by modifying named entities, variable names, narrative framing, and I/O examples while preserving the formal specification and algorithmic structure.
This rewriting acts as a controlled intervention: it alters the surface form through which a model might recognize a problem while holding fixed the underlying algorithmic task and the space of valid solutions.
Rewrites are drafted by GPT-o1 and human-verified for semantic equivalence; we additionally re-run the full test suite to confirm that solutions passing the originals still pass the rewritten versions (\cref{app:annotation_details}).
We then generate solution candidates in five programming languages (Python, C++, Java, Go, Ruby) using two frontier LLMs (Gemini~3~Flash and Claude Sonnet 4.5) and execute each against the full rewritten test suite, recording pass/fail as metadata (\cref{app:solution_gen}).
All candidates are retained regardless of correctness: 3{,}414 across both releases, of which 1{,}065 pass every test case (\emph{verified-correct}).

To test whether rewriting suppresses memorization, we run a controlled $2{\times}2$ study: the same two models are each evaluated on both original and rewritten problems across two releases (\cref{fig:annotation_pass1}).
Because the original and rewritten versions encode the same formal specification, a drop in Pass@1 isolates sensitivity to the problem's surface realization rather than a change in intrinsic difficulty.
For Gemini, Pass@1 decreases consistently after rewriting on \emph{both} releases and across all five languages, indicating that surface-level recognition inflates apparent solving ability even when the algorithmic task is unchanged.
Claude exhibits a release-dependent pattern: it drops on the older release, whose problems are more likely to overlap with its training data, but shows little change on the newer one (per-language breakdowns for both models in \cref{tab:annotation_pass1}).
These results suggest that data leakage varies across models and releases in ways that are difficult to predict without a controlled test.
By rewriting problem statements while preserving their algorithmic content, we aim to reduce the influence of memorization so that benchmark scores better reflect genuine coding and retrieval ability.

\paragraph{Step 3: Query generation.}
We create multiple task-level query variants per problem with LLM assistance, then human-review each for faithfulness and absence of leakage.
Representative instances, full schemas, and per-subtask examples are in \cref{app:task_examples}.

\paragraph{Step 4: Retrieval instance formulation.}
Each retrieval instance pairs a query with its task-specific corpus and is assigned a three-level graded relevance label.
\textbf{Relevance~2 (positive)}: for text-to-code and code-to-code, verified-correct solutions to the queried problem; for code-to-text, the original problem description corresponding to the queried code.
\textbf{Relevance~1 (hard negative)}: same-problem items that are superficially plausible but incorrect: failed solution candidates for text-to-code and code-to-code, and LLM-generated noisy descriptions for code-to-text.
\textbf{Unjudged (easy negative)}: all remaining corpus items, which carry no explicit label.

For all metrics we set \texttt{relevance\_level=2}: binary metrics (Recall, MRR) count only \texttt{relevance=2} items as relevant, and for nDCG hard negatives (\texttt{relevance=1}) are zeroed so they contribute zero gain yet still penalize by occupying top ranks that true positives could occupy.
This formulation is deliberately stricter than CoIR: retrieving a same-problem but incorrect item is not treated as success.
The benchmark can therefore distinguish models that merely recognize the associated problem from those that correctly rank solutions above plausible distractors.

\paragraph{Step 5: Dataset format.}
\textsc{CoREB} is distributed as JSONL files for queries, corpora (code and text), and relevance judgments.
Full field schemas and examples are in \cref{app:coreb_format,app:task_examples}.


\section{Experiments}
\label{section:experiments}

We evaluate \textsc{CoREB} as a two-stage code search benchmark: a first-stage retrieval followed by a reranking stage.
This mirrors the retrieve-then-rerank pipeline used in production code search systems and lets us assess both stages independently.

\subsection{Experimental Setup}
\paragraph{Retrieval models.}
We evaluate eleven embedding models spanning 0.5\,B--8\,B parameters:
\begin{itemize}[nosep,leftmargin=1em]
  \item C2LLM (0.5B, 7B)~\citep{c2llm2025}: code-specialized, adaptive multi-head attention pooling.
  \item F2LLM (0.6B, 1.7B, 4B)~\citep{f2llm2025}: code-specialized, fine-tuned on large datasets.
  \item Jina-code-embeddings (0.5B, 1.5B)~\citep{kryvosheieva2025jinacodeembeddings}: code-specialized embedding models.
  \item Qwen3-Embedding (0.6B, 4B, 8B)~\citep{qwen3embedding2025}: general-purpose encoders.
  \item GemEmb-2~\citep{google2026geminiembedding2}: closed API (Gemini Embedding~2 preview); nominal 3\,B.
\end{itemize}

\paragraph{Reranker baselines.}
For the reranking stage, we compare against four publicly released rerankers, against which we benchmark our fine-tuned \textsc{CoREB-Reranker} (\cref{sec:reranker}):
\begin{itemize}[nosep,leftmargin=1em]
  \item Jina Reranker v2 (base multilingual) and v3~\citep{sturua2024jinarerankerv3}: general-purpose rerankers.
  \item Qwen3-Reranker (0.6B, 4B)~\citep{qwen3embedding2025}: instruction-tuned rerankers, also serve as the backbones for our fine-tuned variants.
\end{itemize}

\paragraph{Metrics.}
All subtasks are evaluated with a uniform ranked-retrieval protocol: each model encodes the query, ranks all corpus items by cosine similarity, and is scored on the resulting ranked list.
We report Recall@$k$ and normalized discounted cumulative gain (nDCG@$k$) in the main text, and include mean reciprocal rank (MRR) in the appendix. Formal definitions are in \cref{app:metrics}.

\begin{table*}[tb]
  \centering
  \small
  \renewcommand{\arraystretch}{1.12}
  \setlength{\tabcolsep}{3.5pt}
  \begin{sc}
  \begin{tabularx}{\textwidth}{l *{8}{S}}
    \toprule
    Model
      & \multicolumn{2}{c}{Text-to-Code}
      & \multicolumn{2}{c}{Code-to-Text}
      & \multicolumn{2}{c}{Code-to-Code$^\dagger$}
      & \multicolumn{2}{c}{Overall} \\
    \cmidrule(lr){2-3} \cmidrule(lr){4-5} \cmidrule(lr){6-7} \cmidrule(lr){8-9}
      & \multicolumn{1}{c}{nDCG} & \multicolumn{1}{c}{Recall}
      & \multicolumn{1}{c}{nDCG} & \multicolumn{1}{c}{Recall}
      & \multicolumn{1}{c}{nDCG} & \multicolumn{1}{c}{Recall}
      & \multicolumn{1}{c}{nDCG} & \multicolumn{1}{c}{Recall} \\
    \midrule
    \rowcolor{aigreen}
    GemEmb-2                  & 0.434 & 0.749 & \textbf{0.813} & \textbf{0.842} & \textbf{0.698} & \textbf{1.000} & \textbf{0.637} & \textbf{0.819} \\
    C2LLM-7B                  & \textbf{0.443} & \textbf{0.753} & 0.795 & \textbf{0.842} & 0.659 & 0.997 & 0.629 & \textbf{0.820} \\
    C2LLM-0.5B                & 0.430 & 0.716 & 0.753 & 0.840          & 0.656 & 0.970 & 0.603 & 0.800 \\
    Jina-code-emb-1.5b        & 0.414 & 0.705 & 0.763 & 0.835          & 0.671 & 0.973 & 0.603 & 0.794 \\
    Jina-code-emb-0.5b        & 0.386 & 0.650 & 0.755 & 0.822          & 0.677 & 0.963 & 0.588 & 0.763 \\
    F2LLM-4B                  & 0.407 & 0.695 & 0.763 & 0.837          & 0.500 & 0.766 & 0.581 & 0.768 \\
    Qwen3-Emb-4B              & 0.390 & 0.626 & 0.728 & 0.828          & 0.392 & 0.603 & 0.546 & 0.717 \\
    F2LLM-1.7B                & 0.383 & 0.603 & 0.715 & 0.805          & 0.383 & 0.562 & 0.536 & 0.692 \\
    F2LLM-0.6B                & 0.344 & 0.545 & 0.665 & 0.793          & 0.334 & 0.491 & 0.491 & 0.654 \\
    Qwen3-Emb-8B              & 0.328 & 0.521 & 0.660 & 0.780          & 0.320 & 0.450 & 0.481 & 0.633 \\
    Qwen3-Emb-0.6B            & 0.349 & 0.541 & 0.617 & 0.755          & 0.384 & 0.551 & 0.477 & 0.641 \\

    \bottomrule
  \end{tabularx}
  \end{sc}
  \vspace{2pt}
   \caption{First-stage retrieval on \textsc{CoREB} \texttt{v202603} (graded qrels, relevance\_level=2). Columns ``nDCG'' and ``Recall'' denote nDCG@10 and Recall@10. Overall is query-count-weighted. Hard negatives (rel=1) penalize nDCG but do not count toward Recall. \textbf{Bold} marks the per-column best.}
  \label{tab:per_task_res_table}
\end{table*}


\subsection{Retrieval Results and Analysis}
\label{sec:analysis}

All analyses in this section are based on \texttt{v202603}; \texttt{v202602} results are in \cref{app:per_release_results}.

\paragraph{Main results.}
\cref{tab:per_task_res_table} reports per-task and overall nDCG@10 and Recall@10 for all eleven models.
GemEmb-2 achieves the highest overall nDCG@10 and leads on code-to-text and code-to-code, while C2LLM-7B is the strongest open-weight model and leads on text-to-code.
Across the board, code-specialized open models (C2LLM, Jina-code) outperform general-purpose encoders of comparable or larger size.
Crucially, no single model dominates all three tasks, confirming that text-to-code, code-to-text, and code-to-code probe complementary capabilities.
Rankings are highly consistent across releases: four of the top-5 models are shared across \texttt{v202602} and \texttt{v202603}, with per-task nDCG@10 differences within 0.03 for every model (\cref{tab:v202602_results}).

\begin{figure}[H]
\noindent
\begin{minipage}[b]{0.45\textwidth}
    \centering
    \includegraphics[width=0.95\linewidth]{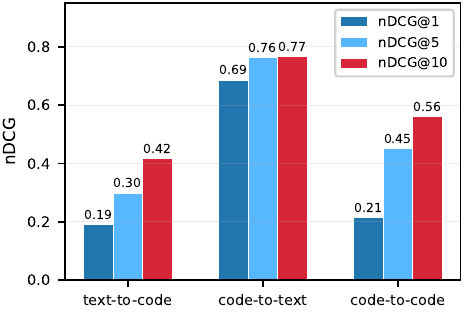}%
    \captionof{figure}{nDCG@$k$ at $k\!\in\!\{1,5,10\}$ per task, averaged over all eleven models on \texttt{v202603}. Code-to-text saturates early; text-to-code and code-to-code grow more steeply with $k$.}
    \label{fig:ndcg_compare}
\end{minipage}%
\hfill
\begin{minipage}[b]{0.49\textwidth}
    \centering
    \small
    \renewcommand{\arraystretch}{0.95}
    \begin{sc}
    \begin{tabular}{lccc}
      \toprule
      Model & Canonical & Full & Search \\
      \midrule
      \rowcolor{aigreen}
      GemEmb-2       & \textbf{0.573} & \textbf{0.565} & 0.000 \\
      C2LLM-7B       & 0.582 & 0.578 & 0.004 \\
      C2LLM-0.5B     & 0.566 & 0.560 & 0.003 \\
      F2LLM-4B       & 0.529 & 0.535 & 0.007 \\
      Jina-code-1.5B  & 0.539 & 0.543 & 0.006 \\
      F2LLM-1.7B     & 0.481 & 0.520 & 0.008 \\
      Qwen3-4B        & 0.510 & 0.504 & \textbf{0.015} \\
      Qwen3-0.6B      & 0.452 & 0.460 & 0.008 \\
      F2LLM-0.6B     & 0.450 & 0.452 & 0.000 \\
      Qwen3-8B        & 0.477 & 0.381 & 0.005 \\
      Jina-code-0.5B  & 0.497 & 0.514 & 0.003 \\
      \bottomrule
    \end{tabular}
    \end{sc}
    \vspace{2pt}
    \captionof{table}{Text-to-code nDCG@10 by subtask (\texttt{v202603}). Canonical and Full use long queries; Search uses short keyword queries ($\sim$19 tokens). Every model collapses on Search.}
    \label{tab:subtask_breakdown}
\end{minipage}
\end{figure}

\paragraph{Analysis I: How much do the three tasks differ in difficulty?}
The three tasks span a wide difficulty range (\cref{fig:ndcg_compare}).
Code-to-text is the easiest (model-averaged nDCG@10 of 0.73), text-to-code is the hardest (0.39), and code-to-code sits in between (0.52) yet is the most discriminative, with a cross-model spread twice that of code-to-text.
The Qwen3 family illustrates this divergence starkly: it scores competitively on code-to-text yet collapses on code-to-code, showing that cross-modal alignment does not transfer to cross-lingual code retrieval.
A single-task benchmark would miss these complementary failure modes entirely.

\paragraph{Analysis II: Does scaling up consistently improve retrieval?}
Larger models do not reliably outperform smaller ones within the same family.
Qwen3-Emb-8B slightly trails Qwen3-Emb-0.6B overall, and drops noticeably on code-to-code (0.320 vs 0.384), the opposite of what a simple scaling law would predict.
Similarly, F2LLM-1.7B underperforms F2LLM-4B despite having triple the parameters.
These non-monotonic curves suggest that training-data composition mediates the size--quality relationship: scaling up a general-purpose encoder does not guarantee better code retrieval.

\paragraph{Analysis III: Can small specialized models compete with large general ones?}
Among open checkpoints, Jina-code-emb-0.5b attains the second-highest code-to-code nDCG@10, trailing only the closed-API GemEmb-2 and edging C2LLM-7B despite being 14$\times$ smaller.
More broadly, C2LLM-0.5B outperforms the 16$\times$ larger Qwen3-Emb-8B by over 12 points overall.
The pattern is consistent: whenever a code-specialized variant exists alongside a general-purpose model of comparable or larger size, the specialized model wins on code-heavy tasks.
For code retrieval, domain-aligned training data matters more than raw model capacity, and a 0.5\,B specialized checkpoint can be a better practical choice than an 8\,B general-purpose one.

\begin{figure}[H]
\noindent
\begin{minipage}{0.48\textwidth}
    \centering
    \includegraphics[width=\linewidth]{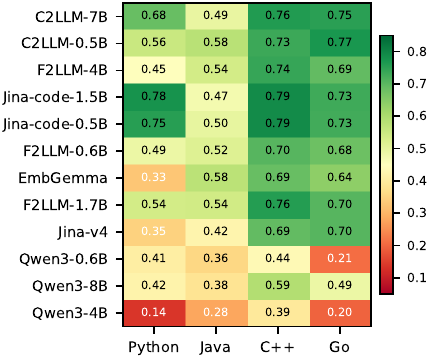}%
    \captionof{figure}{Code-to-code nDCG@10 by anchor language on \texttt{v202603} (after anchor exclusion).}
    \label{fig:c2c_heatmap}
\end{minipage}
\hfill
\begin{minipage}{0.48\textwidth}
    \centering
    \includegraphics[width=\linewidth]{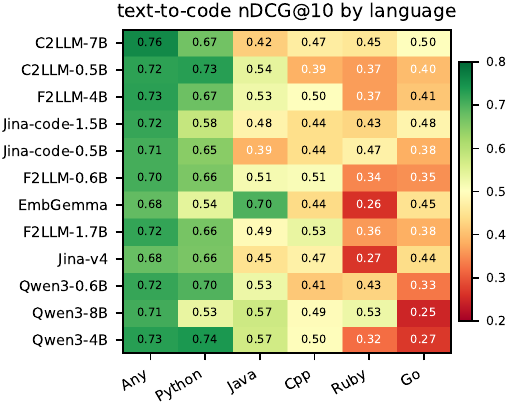}%
    \captionof{figure}{Text-to-code nDCG@10 by target language on \texttt{v202603} (excluding Search subtask).}
    \label{fig:lang_gradient}
\end{minipage}
\end{figure}

\paragraph{Analysis IV: Training data, language, and query length shape retrieval more than model size.}
Beyond the scaling and specialization patterns above, three finer-grained factors emerge from per-subtask and per-language breakdowns (\cref{fig:c2c_heatmap,fig:lang_gradient,tab:subtask_breakdown}):

\noindent\circone \emph{Code-to-code exposes training regime.}
GemEmb-2 and the Jina-code models lead code-to-code by a wide margin, while Qwen3 models collapse to roughly half their nDCG@10 despite competitive text-to-code and code-to-text scores.
Cross-language code-pair training drives this gap; even scaling Qwen3 from 0.6\,B to 8\,B does not close it with the far smaller Jina-code-emb-0.5b.
Per-language breakdowns (\cref{tab:c2c_per_lang}) show C++ and Go anchors are easier to match than Java and Python, consistent with the greater idiomatic diversity of the latter two languages~\citep{zhang2025crosslingual}.

\noindent\circtwo \emph{Short queries collapse all models.}
On the text-to-code Search subtask (19-token keyword queries), every model drops to near-zero nDCG@10, two orders of magnitude below the long-query Canonical subtask (\cref{tab:subtask_breakdown}).
Current retrievers have saturated in the long-query regime while the short-query regime, closest to real search, remains unsolved (see \cref{sec:guidance} for potential mitigations).

\noindent\circthree \emph{Target language introduces systematic bias.}
Language-agnostic text-to-code queries score substantially higher than language-constrained ones (\cref{fig:lang_gradient}), with Ruby and Go lagging consistently across all models.
The gradient tracks training-corpus coverage: Python and Java dominate public code, so models embed them more faithfully.

\noindent Full per-subtask and per-language tables are in \cref{app:subtask_language_analysis}.


\paragraph{Analysis V: Do hard negatives provide additional evaluation signal?}
A key design choice in \textsc{CoREB} is the explicit inclusion of same-problem hard negatives (\texttt{relevance=1}) in the qrels.
Unlike benchmarks with binary or absent-is-irrelevant qrels, our graded scheme exposes \emph{fine-grained discriminability}: a model that ranks a failed code solution or a noise description above the true positive is penalized, even if it places that true positive within the top~k.

To quantify this, we measure the \emph{hard-negative intrusion rate}: among queries where both a positive and a hard negative appear in the top~10, what fraction have at least one hard negative ranked above the highest positive?
\cref{fig:hard_neg_intrusion} shows the result.
On text-to-code, the intrusion rate exceeds 55\% for every model---more than half of all queries have a failed code solution outranking the correct one.
Notably, stronger models (e.g., GemEmb-2, 64\%) can have \emph{higher} intrusion than weaker ones because they retrieve more same-problem content overall (2.1 hard negatives per query vs.\ 1.3 for Qwen3-8B), creating more opportunities for mis-ranking; the metric is conditioned on queries where both a positive and a hard negative reach the top~10.
On code-to-code, intrusion ranges from 43\% (GemEmb-2) to 59\% (F2LLM-4B), with code-specialized models consistently lower than general-purpose ones.
Code-to-text intrusion is much lower (6--30\%), confirming that text retrieval is easier but still exposing a 5$\times$ gap between the best and worst models.
This failure mode, retrieving plausible but incorrect items above true positives, is invisible to benchmarks without explicit hard negatives, yet directly affects deployment quality.

\vspace{6pt}
\noindent
\begin{minipage}{\textwidth}
  \centering
  \includegraphics[width=\linewidth]{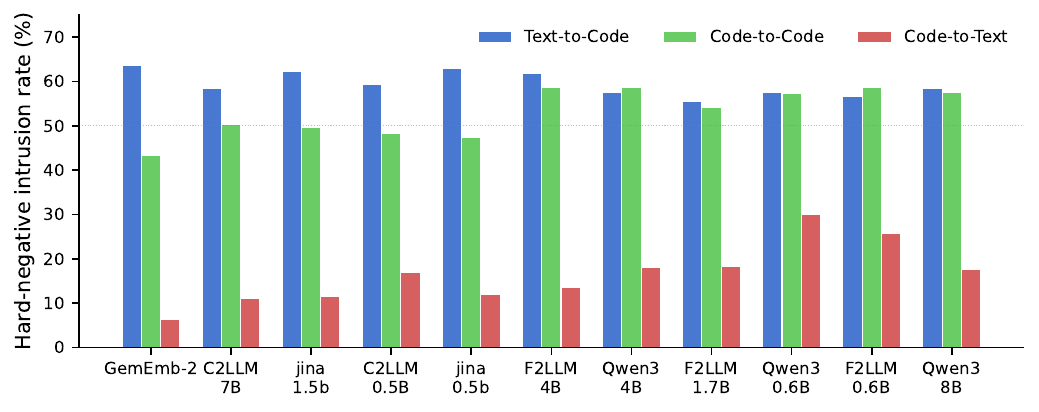}%
  \captionof{figure}{Hard-negative intrusion rate: fraction of queries where at least one hard negative ranks above the best true positive in the top~10. Higher means worse discrimination. Models sorted by overall nDCG@10.}
  \label{fig:hard_neg_intrusion}
\end{minipage}

\subsection{Reranker Evaluation}
\label{sec:reranker}

\begin{wrapfigure}{r}{0.48\textwidth}
  \centering
  \vspace{-12pt}
  \includegraphics[width=\linewidth]{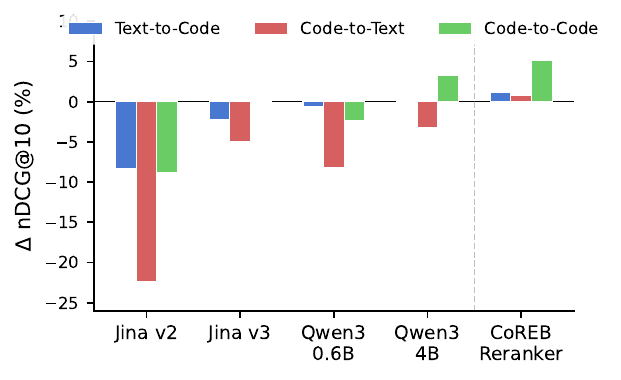}%
  \caption{$\Delta$\,nDCG@10 (\%) after reranking ($k\!=\!128$) on top of C2LLM-7B. No baseline is net-positive across all three tasks; only our fine-tuned \textsc{CoREB-Reranker} achieves this.}
  \label{fig:reranking_delta}
  \vspace{-8pt}
\end{wrapfigure}

\paragraph{Main results on reranking.}
We rerank the top-128 candidates retrieved by C2LLM-7B (the strongest open-weight retriever) with four baseline rerankers: Jina Reranker~v2, Jina Reranker~v3, Qwen3-Reranker-0.6B, and Qwen3-Reranker-4B.
\cref{fig:reranking_delta} reports the nDCG@10 delta per task. No baseline is net-positive across all three tasks.
\circone \emph{Code-to-text}: all four baselines degrade performance (from $-$3.2\% for Qwen3-4B to $-$22.4\% for Jina~v2), because the retriever already saturates near 0.8 nDCG@10 on the compact text corpus and reranking amplifies noise.
\circtwo \emph{Text-to-code}: deltas range from $-$8.3\% (Jina~v2) to $-$0.1\% (Qwen3-4B); no baseline improves this task.
\circthree \emph{Code-to-code}: the only task where reranking helps, with Qwen3-4B gaining $+$3.3\%, as cross-language disambiguation benefits from fine-grained pairwise scoring; however, the other three baselines still hurt.
Overall, a 12-point swing separates the best and worst baseline on the same task, showing that reranker selection matters as much as the decision to rerank itself.

\paragraph{Fine-tuned \textsc{CoREB-Reranker}.}
To mitigate task asymmetry limitations observed in baseline models, we fine-tune Qwen3-Reranker-4B~\citep{qwen3embedding2025} via LoRA~\citep{2022LoRA} on a 3.1M-sample corpus, merging \textsc{CoREB} \texttt{v202602} with datasets including CodeSearchNet~\citep{2019CodeSearchNet,2025CoIR,2021CodeXGLUE}, APPS~\citep{2021APPS}, CosQA~\citep{2021CosQA}, and CodeFeedback~\citep{2024OpenCodeInterpreter}. Training instances are formatted as triplets $(q, d, y)$. We ensure data balance by doubling positive samples and pairing each with one easy and one hard negative. Problem-level disjointness between \texttt{v202602} and the \texttt{v202603} test set is maintained. The released checkpoint is the uniform model soup~\citep{wortsman2022modelsoups} of two LoRA-fine-tuned variants trained from the same initialization with different seeds and data shuffles. (See \cref{app:reranker_training}.)

As shown in \cref{fig:reranking_delta}, \textsc{CoREB-Reranker} is the only reranker in our evaluation that is net-positive across all three tasks, establishing a complete two-stage pipeline, retrieval (C2LLM-7B) followed by in-domain reranking, that improves over the retriever alone on every task.


\vspace{6pt}
\noindent
\begin{minipage}{0.48\textwidth}
  \centering
  \includegraphics[width=\linewidth]{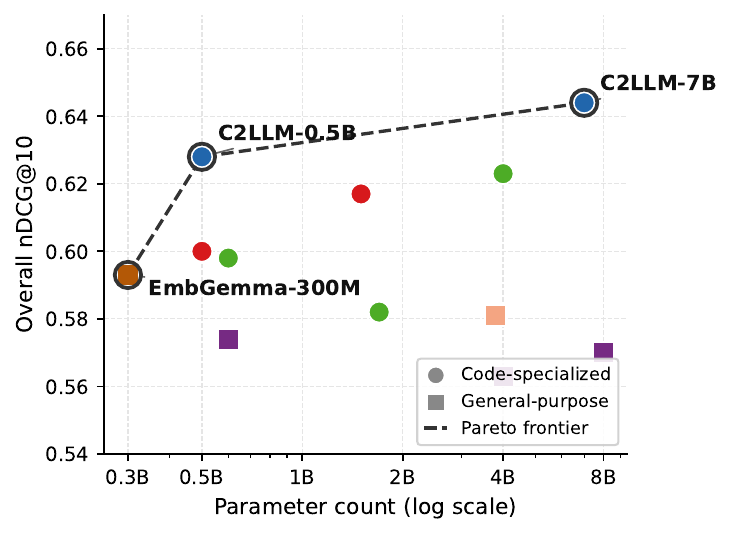}%
  \captionof{figure}{Overall nDCG@10 vs.\ parameter count (log scale) for the ten open-weight models. Circles = code-specialized; squares = general-purpose. The dashed line marks the Pareto frontier. GemEmb-2 is excluded due to its unknown parameter size.}
  \label{fig:efficiency_scatter}
\end{minipage}
\hfill
\begin{minipage}{0.48\textwidth}
  \centering
  \includegraphics[width=\linewidth]{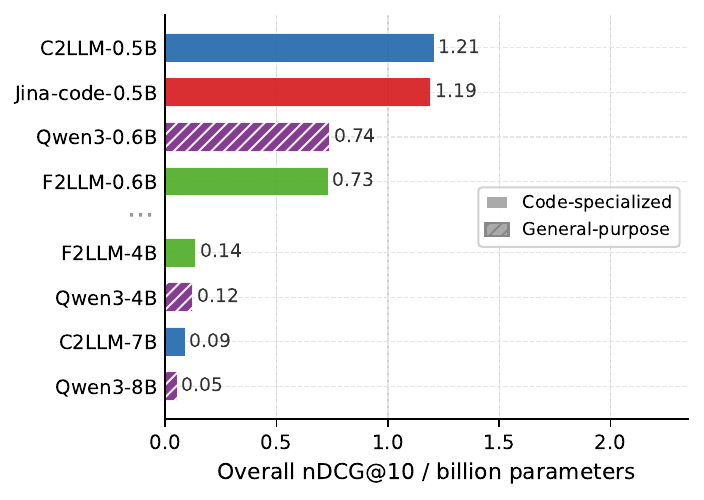}%
  \captionof{figure}{Parameter efficiency (nDCG@10 per billion parameters) for the four most and four least efficient models; middle-ranked models are omitted (full ranking in \cref{fig:efficiency_full}).}
  \label{fig:efficiency_bar}
\end{minipage}

\subsection{Practical Guidance for Code Search}
\label{sec:guidance}

Our analyses yield concrete recommendations for practitioners building code search systems.

\paragraph{Choose code-specialized embeddings.}
\cref{fig:efficiency_scatter} plots overall nDCG@10 against parameter count for all eleven models.
Among open-weight checkpoints, two lie on the Pareto frontier: C2LLM-0.5B and C2LLM-7B; every other open model is simultaneously outperformed and outscaled by a smaller open checkpoint.
C2LLM-0.5B reaches 95.9\% of the best open-weight score at only 7\% of the parameter count, and 0.5\,B code-specialized models deliver roughly 13$\times$ more nDCG per billion parameters than 8\,B general-purpose ones (\cref{fig:efficiency_bar}).
For latency- or memory-constrained deployments, a 0.5\,B code-specialized checkpoint is the clear choice at any scale evaluated here.

\paragraph{Reranker selection is high-stakes.}
Off-the-shelf rerankers are task-asymmetric: a 12-point swing separates the best and worst baseline on the same task (\cref{fig:reranking_delta}).
A poorly chosen reranker can degrade code-to-text by over 20 points.
In-domain fine-tuning on graded qrels (as in our \textsc{CoREB-Reranker}) is needed to achieve consistent gains across all three retrieval directions.

\paragraph{Short-query retrieval remains an open problem.}
The near-zero performance on keyword-style Search queries (\cref{tab:subtask_breakdown}) is the single largest unsolved gap.
Neither scaling the embedding model nor adding a reranker closes it; query expansion techniques such as HyDE~\citep{gao2023hyde} or Query2doc~\citep{wang2023query2doc} are the most promising path forward.

\section{Conclusion}
We presented \textsc{CoREB}, a contamination-limited benchmark and reranker that covers both stages of the code search pipeline across three tasks and five languages.
Our experiments show that code-specialised training matters more than scale, that short keyword queries and low-resource languages remain unsolved failure modes, and that reranker selection is high-stakes.

\clearpage

\bibliographystyle{icml2020_url}
\bibliography{reference}

\clearpage
\appendix
\appendixpage

\crefalias{section}{appendix}
\crefalias{subsection}{appendix}

\section{Dataset Details}
\label{app:dataset}

\subsection{Annotation Details}
\label{app:annotation_details}

\paragraph{Code Generation.}
Each model receives a fixed prompt that specifies the target programming
language, problem title, and full problem statement.
Template variables are written as \texttt{\{\{~\}\}}.

\begin{lstlisting}[caption={Prompt template for code generation.},
                   label={lst:code_gen_prompt}]
You are an expert competitive-programming assistant.

Task
----
Write an *entire*, *runnable* program in {{language}} that solves
the problem below.

  Title    : {{question_title}}
  Statement: {{question_content}}

Requirements
------------
1. Provide ONLY executable source code -- no comments, no markdown,
   no explanations.
2. Implement a `main()` function that:
     - reads all input from stdin,
     - computes the answer,
     - writes the result to stdout.
3. Add any necessary helper functions (without comments).
4. Call `main()` at the bottom of the file.
5. Do not modify the supplied starter skeleton:

   {{starter_code}}

Output Format
-------------
Wrap the final program in XML-style tags:

  <code>
  ...your code...
  </code>

Generate the COMPLETE solution. Do not stop mid-function.
\end{lstlisting}

\paragraph{Counterfactual Rewriting.}
To reduce surface-level contamination, each problem is passed through
an LLM that applies five transformations while preserving the
underlying algorithmic challenge.  Purely numerical test cases are
never modified; non-numerical test cases receive only the minimal
changes required to match the new surface context.

\begin{lstlisting}[caption={Simplified prompt template for counterfactual problem rewriting.
The actual implementation includes Jinja2 template conditionals for handling
optional test case fields.},
                   label={lst:annotate_prompt}]
# Code Problem Annotation Task

## Task Description
You are tasked with transforming a coding problem into a counterfactual
version to avoid data contamination. Your goal is to preserve the core
algorithmic challenge without changing the output while changing superficial
details.

## Transformation Instructions
Please create a counterfactual version of this problem by applying these
transformations:

1. **Named Entity Replacement**:
   - Replace all proper nouns, character names, company names, etc.
   - Example: "Alice wants to sort her books"
     -> "Marcus needs to organize his collection"

2. **Domain/Context Shifting**:
   - Change the problem domain while keeping the algorithmic challenge
     identical
   - Example: "Calculate profit from stock trades"
     -> "Determine score changes in a game tournament"

3. **Noun Phrase Substitution**:
   - Replace key objects/items with different but functionally equivalent
     ones
   - Example: "array of integers" can stay the same, but
     "list of books" -> "array of products"

4. **Synonym Replacement**:
   - Replace verbs and adjectives with synonyms
   - Example: "maximize profit" -> "optimize earnings"

5. **Variable/Function Name Changes**:
   - If example code is provided, rename variables and functions
   - Example: `calculateSum()` -> `computeTotal()`

## Important Guidelines
- Preserve the exact same algorithmic challenge and difficulty
- Maintain the same input/output structure and constraints
- Keep the same time/space complexity requirements
- Ensure the transformed problem requires the same solution approach
- The starter code should remain syntactically correct and functionally
  equivalent

## Test Case Guidelines
When handling test cases, follow these strict rules:

1. For purely numerical test cases (containing only numbers, basic
   operators, and data structures):
   - DO NOT MODIFY them at all - keep them exactly as they are
   - Example: Leave "[1, 2, 3] -> 6" or "5 + 10 = 15" unchanged

2. For non-numerical test cases (containing domain-specific terms):
   - Make MINIMAL changes necessary to match your transformed problem
     context
   - PRESERVE the exact same algorithmic structure and complexity
   - Maintain the same input/output patterns and edge cases
   - Example: If you changed "count books on shelf" to "count tools in
     box", then "books=['novel','textbook'] -> 2" becomes
     "tools=['hammer','wrench'] -> 2"

3. ALL test cases must:
   - Remain syntactically correct in the target language
   - Test exactly the same edge cases and functionality
   - Have the same expected outputs for equivalent inputs

## Your Counterfactual Version

Given the original problem:
Title: {{question_title}}

Content:
{{question_content}}

Starter Code:
{{starter_code}}

Public Test Cases:
{{public_test_cases}}

Private Test Cases:
{{private_test_cases}}

Please provide your transformed version in JSON format with this structure:

{
  "annotate_question_title": "<transformed_title>",
  "annotate_question_content": "<transformed_content>",
  "annotate_starter_code": "<transformed_starter_code>",
  "annotate_public_test_cases": "<transformed_public_test_cases>",
  "annotate_private_test_cases": "<transformed_private_test_cases>"
}

Note: Ensure that your transformed version preserves all the algorithmic
details while changing the superficial context. The code should remain
valid and compilable.
\end{lstlisting}

\paragraph{Question Abbreviation.}
To support the Canonical Retro retrieval sub-tasks, each
(rewritten) problem description is further condensed into a
retrieval-optimized summary of 50--150 words that retains only
the core goal, key constraints, and distinctive terminology.

\begin{lstlisting}[caption={Prompt template for question abbreviation.},
                   label={lst:abbrev_prompt}]
# Question Content Abbreviation Task

## Task Description
You are an expert at creating concise, retrieval-optimized summaries of
coding problem descriptions. Your goal is to distill the essential
information from a problem statement into a compact format that preserves
key details while removing redundancy.

## Input
Question title: {{question_title}}
Content: {{question_content}}

## Core Principles
**Essential Information Preservation**
- Retain the core problem goal, key constraints, and unique requirements
- Preserve specific numerical limits, data types, and optimization
  objectives
- Keep distinctive algorithmic concepts or problem-specific terminology

**Conciseness Optimization**
- Remove verbose explanations, examples, and repetitive content
- Eliminate unnecessary background information or motivational text
- Compress similar concepts into unified statements

**Retrieval-Friendly Format**
- Use clear, structured language that matches how developers search
- Include specific technical terms and constraints that distinguish this
  problem
- Maintain logical flow from problem statement to requirements

## Abbreviation Guidelines
1. **Problem Goal**: Start with a clear, concise statement of what needs
   to be accomplished
2. **Key Constraints**: Include specific limits, ranges, and requirements
   (e.g., "array size <= 10^5", "time complexity O(n log n)")
3. **Input/Output Format**: Specify data types and formats when critical
   to the problem
4. **Unique Requirements**: Highlight distinctive aspects that
   differentiate this problem from similar ones
5. **Algorithmic Hints**: Include key algorithmic concepts if they're
   central to the solution approach

## Output Requirements
- **Length**: 50-150 words maximum
- **Structure**: Use clear, declarative sentences
- **Precision**: Include specific numerical constraints and technical
  details
- **Clarity**: Avoid ambiguous pronouns or references
- **Completeness**: Ensure all essential problem-solving information is
  preserved

Please provide your abbreviated version in JSON format:

{
  "abbreviated_content": "<your_concise_abbreviation>"
}
\end{lstlisting}

\subsection{Hard Negative Generation}
\label{app:noise_gen}

To strengthen text-retrieval evaluation, we augment each release's original problem descriptions with four LLM-generated hard negatives per problem, yielding 668 hard negatives in \texttt{v202602} and 700 in \texttt{v202603} (1{,}368 aggregated across releases).
A \emph{hard negative} is a problem description that shares surface-level vocabulary or
structural similarity with the original but differs in its core algorithmic challenge, making it
difficult for a retrieval model to distinguish from the true positive without deep semantic
understanding.

Hard negatives are generated by prompting Qwen-32B with one of four perturbation strategies,
applied independently and with varied temperature ($\in [0.85, 1.0]$) to encourage diversity:

\begin{enumerate}[leftmargin=*,itemsep=0.2em]
  \item Operation-type change.  Alter the fundamental operation the algorithm must
        perform (e.g., replace a ``remove / delete'' goal with a ``select / construct / merge''
        goal) so that the core algorithmic action is different, not merely renamed.

  \item Optimization-objective change.  Invert or replace the optimization criterion
        (e.g., change ``maximize'' to ``minimize'' or ``count distinct'') so that what the
        algorithm optimizes for changes structurally.

  \item Algorithmic-approach change.  Replace the algorithmic paradigm required to solve
        the problem (e.g., subsequence reasoning $\to$ contiguous-array or graph problems;
        greedy $\to$ dynamic programming; two-pointer $\to$ binary search) so that the
        required solution strategy is qualitatively different.

  \item Problem-domain change.  Alter input data types and the problem context
        (e.g., strings $\to$ graphs, arrays $\to$ trees), producing a structurally distinct problem
        that shares no obvious surface mapping to the original.
\end{enumerate}

\noindent Generated texts are post-processed with a regular-expression pass to strip LLM-produced
markdown headers and formatting artifacts (e.g., ``\textbf{Modified Problem Description:}'')
before being added to the corpus.
Each release's text corpus (835 entries in \texttt{v202602} and 875 in \texttt{v202603}; 1{,}710 aggregated) serves as the retrieval corpus for its code-to-text and text-to-code subtasks; see \cref{tab:corpus_composition} for the per-release and aggregated composition.

\paragraph{Code hard negatives in qrels (v2 scheme).}
In addition to LLM-generated text hard negatives, the v2 qrel scheme also encodes \emph{code} hard negatives directly in the relevance judgments.
For text-to-code and code-to-code, same-problem code solutions that \emph{failed} execution tests are assigned \texttt{relevance=1}; when the failed pool is insufficient, correct solutions that do not qualify as positives under the current subtask constraints (e.g., wrong language for a language-specific query) are used instead.
For code-to-text, the four LLM-generated noise descriptions per problem are assigned \texttt{relevance=1}, including texts from the same problem (highest confusability) and texts from unrelated problems as a fallback.
These explicitly judged hard negatives allow a retrieval model's score to be penalized when it ranks a plausible-but-incorrect document above a true positive, a failure mode that absent-is-irrelevant schemes cannot detect.

\begin{table}[t]
  \centering
  \small
  \renewcommand{\arraystretch}{1.15}
  \setlength{\tabcolsep}{4pt}
  \begin{sc}
  \begin{tabular}{l l r r r}
    \toprule
    \textbf{Task} & \textbf{Rel.} & \textbf{v202602} & \textbf{v202603} & \textbf{Source} \\
    \midrule
    \multirow{2}{*}{T2C}
      & 2 (pos)      & 2{,}742 & 2{,}814 & Passed solutions \\
      & 1 (hard neg)  & 3{,}320 & 3{,}136 & Failed solutions\textsuperscript{a} \\
    \midrule
    \multirow{2}{*}{C2T}
      & 2 (pos)      & 1{,}064 & 1{,}010 & Original descriptions \\
      & 1 (hard neg)  & 3{,}810 & 3{,}600 & LLM noise texts\textsuperscript{b} \\
    \midrule
    \multirow{2}{*}{C2C}
      & 2 (pos)      &   367 &   623 & Cross-lang translations \\
      & 1 (hard neg)  &   507 &   834 & Failed / excluded solutions\textsuperscript{c} \\
    \midrule
    \multicolumn{2}{l}{\textbf{Total}}
      & 11{,}810 & 12{,}017 & \\
    \bottomrule
  \end{tabular}
  \end{sc}
  \vspace{2pt}
  {\footnotesize\raggedright
  \textsuperscript{a}Same-problem code that failed execution tests; when insufficient, correct solutions excluded by subtask constraints (e.g., wrong language).
  \textsuperscript{b}Four LLM-generated noise descriptions per problem, prioritizing same-problem texts.
  \textsuperscript{c}Failed solutions for the same problem, supplemented by correct solutions that do not qualify as positives (e.g., same-language pairs in a cross-language subtask).}
  \caption{Graded relevance judgment counts per release under the v2 scheme. \texttt{relevance=2} items are true positives; \texttt{relevance=1} items are hard negatives that penalize nDCG when ranked above true positives. Unjudged corpus items act as easy negatives.}
  \label{tab:graded_qrel_counts}
\end{table}

\subsection{Main Tasks and Subtasks}
\label{app:coreb_task_subtask_defs}

CoREB structures evaluation as a hierarchy of \emph{main tasks} and
\emph{subtasks}. A main task specifies the retrieval direction (i.e., what modality is used as the query and what modality constitutes the retrieval corpus). A subtask instantiates a main task under a particular
configuration (e.g., query style, language constraint, or invariance
setting), enabling controlled comparisons and fine-grained diagnosis.

\paragraph{Main task definition.}
Each query instance belongs to one of three main tasks, recorded in the
\texttt{split} field:
\begin{itemize}[leftmargin=*]
  \item text-to-code: the query is a natural language
        problem statement and the retrieval target is a code implementation.
  \item code-to-code: the query is a code snippet and
        the retrieval target is another code implementation with equivalent
        semantics.
  \item code-to-text: the query is a code implementation
        and the retrieval target is a natural language problem statement.
\end{itemize}
The main task determines the query modality, corpus modality, and
evaluation protocol.

\paragraph{Subtask definition.}
Each query carries a \texttt{subtask} identifier that names the complete
subtask, e.g.\ \texttt{t2c\_canonical\_retro\_python} or
\texttt{c2c\_cross\_lang}. The subtask variant component captures the
evaluation setting (query style, matching regime), while the language
suffix, when present, specifies the language slice for that instantiation.
Query-level metadata fields provide additional context (e.g.,
\texttt{anchor\_language}, \texttt{anchor\_model},
\texttt{sub\_query\_type}).

\paragraph{text-to-code subtasks.}
Three subtask variants are defined:

\begin{itemize}[leftmargin=*, itemsep=0.3em]
  \item \texttt{canonical\_retro}: the query is an \emph{abbreviated}
        problem description retaining only the core goal and key
        constraints, omitting narrative detail
        (\texttt{sub\_query\_type=abbreviated}).

  \item \texttt{full\_retro}: the query is the \emph{full} problem
        statement including contextual details, examples, and constraints
        (\texttt{sub\_query\_type=full\_description}).

  \item \texttt{search}: the query is an LLM-generated developer-style
        search string. The \texttt{sub\_query\_type} field further
        distinguishes intent-focused natural language
        (\texttt{description\_search}), technique-focused queries naming
        specific algorithms or data structures (\texttt{algorithm\_search}),
        and broad language-agnostic queries (\texttt{language\_agnostic}).
\end{itemize}

\noindent\emph{Language slicing.}
Each variant is instantiated for each of the five target languages
(\texttt{language\_constraint} $\in \{\texttt{python},\ \texttt{java},\
\texttt{cpp},\ \texttt{ruby},\ \texttt{go}\}$) and for an unconstrained
variant (suffix \texttt{*\_any}, \texttt{language\_constraint=none}),
yielding complete subtasks such as \texttt{t2c\_canonical\_retro\_python}
and \texttt{t2c\_search\_any}. When a concrete language is specified, only
corpus entries in that language are treated as valid retrieval targets;
otherwise any language is accepted.

\paragraph{code-to-code subtasks.}
The current release contains one code-to-code subtask variant, using code as both query and corpus.
The anchor (query) language and model are recorded in
\texttt{anchor\_language} and \texttt{anchor\_model}; intended target
languages are listed in \texttt{meta.target\_languages}.

\begin{itemize}[leftmargin=*, itemsep=0.3em]
  \item \texttt{cross\_lang}: cross-language semantic alignment. The
        anchor is a solution in one language (Python, Java, C++, or Go)
        generated by Claude~Sonnet~4.5, and the task is to retrieve the
        semantically equivalent solution in a different target language
        by the same model (e.g., Java anchor $\to$ Python target, or
        Python anchor $\to$ Ruby target), testing cross-language
        transfer robustness.
\end{itemize}

\noindent Language is encoded per-query via \texttt{anchor\_language} and
\texttt{meta.target\_languages}; the subtask therefore carries no language suffix.

\paragraph{code-to-text subtasks.}
Three subtask variants are defined:

\begin{itemize}[leftmargin=*, itemsep=0.3em]
  \item \texttt{canonical\_retro}: retrieve \emph{abbreviated} problem
        descriptions from code
        (\texttt{meta.text\_type=canonical\_description}).

  \item \texttt{full\_retro}: retrieve \emph{full} problem descriptions
        from code (\texttt{meta.text\_type=full\_description}).

  \item \texttt{match}: code--text pair retrieval. Each query is a code
        snippet and the target is its corresponding problem description
        among the release's full text corpus (835 entries in \texttt{v202602}, 875 in \texttt{v202603}). Queries carry a binary
        \texttt{label} (1~=~positive pair, 0~=~negative pair) and a
        \texttt{meta.pair\_type} field for analysis; however, evaluation
        uses the same ranked-retrieval protocol as the other subtasks
        (nDCG@10, Recall@10, etc.).  Positive queries have exactly one
        relevant document; negative queries have none and are excluded
        from metric averages following standard IR convention.
\end{itemize}

\noindent\emph{Language slicing.}
For the retrieval variants (\texttt{canonical\_retro},
\texttt{full\_retro}), the anchor language is recorded in
\texttt{anchor\_language}. The language-agnostic variant (\texttt{*\_any})
mixes queries from all anchor languages and is evaluated as standard
description retrieval; language-specific variants fix the anchor language
and probe cross-language invariance, yielding subtasks such as
\texttt{c2t\_full\_retro\_java}. The \texttt{match} subtask is
language-agnostic and carries no language suffix.

\paragraph{Rationale.}
This hierarchy isolates different sources of difficulty (linguistic variability, verbosity, keyword-style queries, and cross-language generalization) while preserving a consistent evaluation interface across tasks. As a result, CoREB supports both overall benchmarking and diagnostic analysis at the subtask level.

\begin{table*}[t]
  \centering
  \small
  \renewcommand{\arraystretch}{1.1}
  \setlength{\tabcolsep}{6pt}
  \begin{sc}
  \begin{tabular}{l l l l r}
    \toprule
    Complete Task Name & Text Type & Query Type & Language & Count \\
    &   &   & Constraint &  \\
    \midrule
    \texttt{t2c\_canonical\_retro\_any}     & abbreviated       & language\_agnostic & any    & 103 \\
    \texttt{t2c\_canonical\_retro\_python}  & abbreviated       & language\_specific & python & 79 \\
    \texttt{t2c\_canonical\_retro\_java}    & abbreviated       & language\_specific & java   & 63 \\
    \texttt{t2c\_canonical\_retro\_cpp}     & abbreviated       & language\_specific & cpp    & 70 \\
    \texttt{t2c\_canonical\_retro\_ruby}    & abbreviated       & language\_specific & ruby   & 73 \\
    \texttt{t2c\_canonical\_retro\_go}      & abbreviated       & language\_specific & go     & 63 \\
    \texttt{t2c\_full\_retro\_any}          & full\_desc & language\_agnostic & any    & 103 \\
    \texttt{t2c\_full\_retro\_python}       & full\_desc & language\_specific & python & 79 \\
    \texttt{t2c\_full\_retro\_java}         & full\_desc & language\_specific & java   & 63 \\
    \texttt{t2c\_full\_retro\_cpp}          & full\_desc & language\_specific & cpp    & 70 \\
    \texttt{t2c\_full\_retro\_ruby}         & full\_desc & language\_specific & ruby   & 73 \\
    \texttt{t2c\_full\_retro\_go}           & full\_desc & language\_specific & go     & 63 \\
    \texttt{t2c\_search\_any}               & search\_mixed     & language\_agnostic & any    & 13 \\
    \texttt{t2c\_search\_python}            & search\_mixed     & language\_specific & python & 50 \\
    \texttt{t2c\_search\_java}              & search\_mixed     & language\_specific & java   & 50 \\
    \texttt{t2c\_search\_cpp}               & search\_mixed     & language\_specific & cpp    & 50 \\
    \texttt{t2c\_search\_ruby}              & search\_mixed     & language\_specific & ruby   & 50 \\
    \texttt{t2c\_search\_go}                & search\_mixed     & language\_specific & go     & 50 \\
    \bottomrule
  \end{tabular}
  \end{sc}
  \caption{Summary of text-to-code subtasks.}
  \label{tab:t2c_subtasks}
\end{table*}

\begin{table*}[t]
  \centering
  \small
  \renewcommand{\arraystretch}{1.1}
  \setlength{\tabcolsep}{3pt}
  \begin{sc}
  \begin{tabular}{l l l l r}
    \toprule
    Complete Task Name & Text Type & Query Type & Anchor & Count \\
    &  &   & Language & \\
    \midrule
    \texttt{c2c\_cross\_lang}  & N/A & cross\_lang  & python & 35 \\
    \texttt{c2c\_cross\_lang}  & N/A & cross\_lang  & java   & 49 \\
    \texttt{c2c\_cross\_lang}  & N/A & cross\_lang  & cpp    & 47 \\
    \texttt{c2c\_cross\_lang}  & N/A & cross\_lang  & go     & 38 \\
    \bottomrule
  \end{tabular}
  \end{sc}
  \caption{Overview of code-to-code subtasks.}
  \label{tab:c2c_subtasks}
\end{table*}

\begin{table*}[t]
  \centering
  \small
  \renewcommand{\arraystretch}{1.1}
  \setlength{\tabcolsep}{4pt}
  \begin{sc}
  \begin{tabular}{l l l l r}
    \toprule
    Complete Task Name & Text Type & Query Type & Anchor  & Count \\
    &    &    & Language & \\
    \midrule
    \texttt{c2t\_canonical\_retro\_any}   & canonical\_desc & description\_retrieval       & python & 79 \\
    \texttt{c2t\_canonical\_retro\_any}   & canonical\_desc & description\_retrieval       & cpp    & 9  \\
    \texttt{c2t\_canonical\_retro\_any}   & canonical\_desc & description\_retrieval       & ruby   & 6  \\
    \texttt{c2t\_canonical\_retro\_any}   & canonical\_desc & description\_retrieval       & go     & 6  \\
    \texttt{c2t\_canonical\_retro\_any}   & canonical\_desc & description\_retrieval       & java   & 3  \\
    \texttt{c2t\_canonical\_retro\_python}& canonical\_desc & cross\_language\_invariance  & python & 67 \\
    \texttt{c2t\_canonical\_retro\_java}  & canonical\_desc & cross\_language\_invariance  & java   & 63 \\
    \texttt{c2t\_canonical\_retro\_cpp}   & canonical\_desc & cross\_language\_invariance  & cpp    & 67 \\
    \texttt{c2t\_canonical\_retro\_ruby}  & canonical\_desc & cross\_language\_invariance  & ruby   & 68 \\
    \texttt{c2t\_canonical\_retro\_go}    & canonical\_desc & cross\_language\_invariance  & go     & 61 \\
    \texttt{c2t\_full\_retro\_any}        & full\_desc      & description\_retrieval       & python & 79 \\
    \texttt{c2t\_full\_retro\_any}        & full\_desc      & description\_retrieval       & cpp    & 9  \\
    \texttt{c2t\_full\_retro\_any}        & full\_desc      & description\_retrieval       & ruby   & 6  \\
    \texttt{c2t\_full\_retro\_any}        & full\_desc      & description\_retrieval       & go     & 6  \\
    \texttt{c2t\_full\_retro\_any}        & full\_desc      & description\_retrieval       & java   & 3  \\
    \texttt{c2t\_full\_retro\_python}     & full\_desc      & cross\_language\_invariance  & python & 67 \\
    \texttt{c2t\_full\_retro\_java}       & full\_desc      & cross\_language\_invariance  & java   & 63 \\
    \texttt{c2t\_full\_retro\_cpp}        & full\_desc      & cross\_language\_invariance  & cpp    & 67 \\
    \texttt{c2t\_full\_retro\_ruby}       & full\_desc      & cross\_language\_invariance  & ruby   & 68 \\
    \texttt{c2t\_full\_retro\_go}         & full\_desc      & cross\_language\_invariance  & go     & 61 \\
    \texttt{c2t\_match}                   & N/A             & pair\_match\_retrieval        & python & 316 \\
    \texttt{c2t\_match}                   & N/A             & pair\_match\_retrieval        & cpp    & 36 \\
    \texttt{c2t\_match}                   & N/A             & pair\_match\_retrieval        & go     & 24 \\
    \texttt{c2t\_match}                   & N/A             & pair\_match\_retrieval        & ruby   & 24 \\
    \texttt{c2t\_match}                   & N/A             & pair\_match\_retrieval        & java   & 12 \\
    \bottomrule
  \end{tabular}
  \end{sc}
  \caption{Summary of code-to-text subtasks.}
  \label{tab:c2t_subtasks}
\end{table*}

\begin{table}[t]
  \centering
  \small
  \renewcommand{\arraystretch}{1.15}
  \setlength{\tabcolsep}{4pt}
  \begin{sc}
  \begin{tabular}{l l r r}
    \toprule
    \textbf{Main Task} & \textbf{Subtask} & \textbf{v202602} & \textbf{v202603} \\
    \midrule
    \multirow{3}{*}{T2C}
      & Canonical Retro & 123 & 117 \\
      & Full Retro      & 448 & 414 \\
      & Search          &  25 &  13 \\
    \midrule
    \multirow{3}{*}{C2C}
      & Cross-Lingual   & 215 & 200 \\
      & Cross-Model     & --- & 182 \\
      & Mono-Lang       & --- & 200 \\
    \midrule
    \multirow{3}{*}{C2T}
      & Canonical Retro & 255 & 238 \\
      & Full Retro      & 255 & 238 \\
      & Pair Match      & 259 & 244 \\
    \bottomrule
  \end{tabular}
  \end{sc}
  \caption{Per-release average query length (tokens, \texttt{cl100k\_base}) by subtask. ``---'' indicates a subtask absent from that release. Cross-Model and Mono-Lang are \texttt{v202603}-only C2C subtasks. Aggregated values are reported in the main text (\cref{tab:query_subtasks}).}
  \label{tab:query_tokens_per_release}
\end{table}

\begin{table}[t]
  \centering
  \small
  \renewcommand{\arraystretch}{1.2}
  \setlength{\tabcolsep}{5pt}
  \begin{sc}
  \begin{tabular}{l l c c c c c c c}
    \toprule
    \textbf{Task} & \textbf{Sub-Task} & \textbf{Any} & \textbf{Py} & \textbf{Java} & \textbf{C++} & \textbf{Ruby} & \textbf{Go} & \textbf{Total} \\
    \midrule
    \multirow{3}{*}{Text-to-Code}
      & Canonical Retro & 103 & 79 & 63 & 70 & 73 & 63 & 451 \\
    & Full Retro        & 103 & 79 & 63 & 70 & 73 & 63 & 451 \\
    & Search            & 13  & 50 & 50 & 50 & 50 & 50 & 263 \\
    \midrule
    \multirow{2}{*}{Code-to-Text}
      & Canonical Retro & 103 & 67 & 63 & 67 & 68 & 61 & 429 \\
    & Full Retro        & 103 & 67 & 63 & 67 & 68 & 61 & 429 \\
    \midrule
    \multicolumn{2}{l}{\textbf{Total}} & 425 & 342 & 302 & 324 & 332 & 298 & \textbf{2,023} \\
    \bottomrule
  \end{tabular}
  \end{sc}
  \caption{Per-language query counts for language-variant sub-tasks in \textsc{CoREB}. \emph{Any} = language-agnostic variant. Variation in counts across languages reflects the available evaluated solutions per problem.}
  \label{tab:language_breakdown}
\end{table}

\begin{table}[t]
  \centering
  \small
  \renewcommand{\arraystretch}{1.15}
  \setlength{\tabcolsep}{5pt}
  \begin{sc}
  \begin{tabular}{l l r r r}
    \toprule
    \textbf{Dimension} & \textbf{Category} & \textbf{v202602} & \textbf{v202603} & \textbf{Aggregated} \\
    \midrule
    \multirow{3}{*}{Main Task}
      & Text-to-Code & 1{,}165 & 1{,}117 & 2{,}282 \\
    & Code-to-Code &    169 &   166 &   335 \\
    & Code-to-Text & 1{,}270 & 1{,}200 & 2{,}470 \\
    \midrule
    \multirow{6}{*}{Language}
      & Any    & 425 & 391 &   816 \\
      & Python & 693 & 635 & 1{,}328 \\
      & Java   & 363 & 317 &   680 \\
      & C++    & 407 & 403 &   810 \\
      & Ruby   & 356 & 344 &   700 \\
      & Go     & 360 & 393 &   753 \\
    \midrule
    \multirow{5}{*}{Subtask Variant}
      & Canonical Retro &  880 &  838 & 1{,}718 \\
      & Full Retro      &  880 &  838 & 1{,}718 \\
      & Search          &  263 &  261 &   524 \\
      & Cross-Lingual   &  169 &  166 &   335 \\
      & Pair Match      &  412 &  380 &   792 \\
    \midrule
    \multicolumn{2}{l}{\textbf{Total}} & \textbf{2{,}604} & \textbf{2{,}483} & \textbf{5{,}087} \\
    \bottomrule
  \end{tabular}
  \end{sc}
  \caption{Detailed query counts for each \textsc{CoREB} release and the aggregated total, broken down by main task, language constraint, and subtask variant. Supplements the high-level corpus overview in \cref{tab:coreb_releases}; visualised in \cref{fig:task_overview,fig:task_overview_per_release}.}
  \label{tab:query_dist_summary}
\end{table}

\begin{figure*}[t]
  \centering
  \begin{subfigure}[b]{0.32\linewidth}
    \centering
    \includegraphics[width=\linewidth]{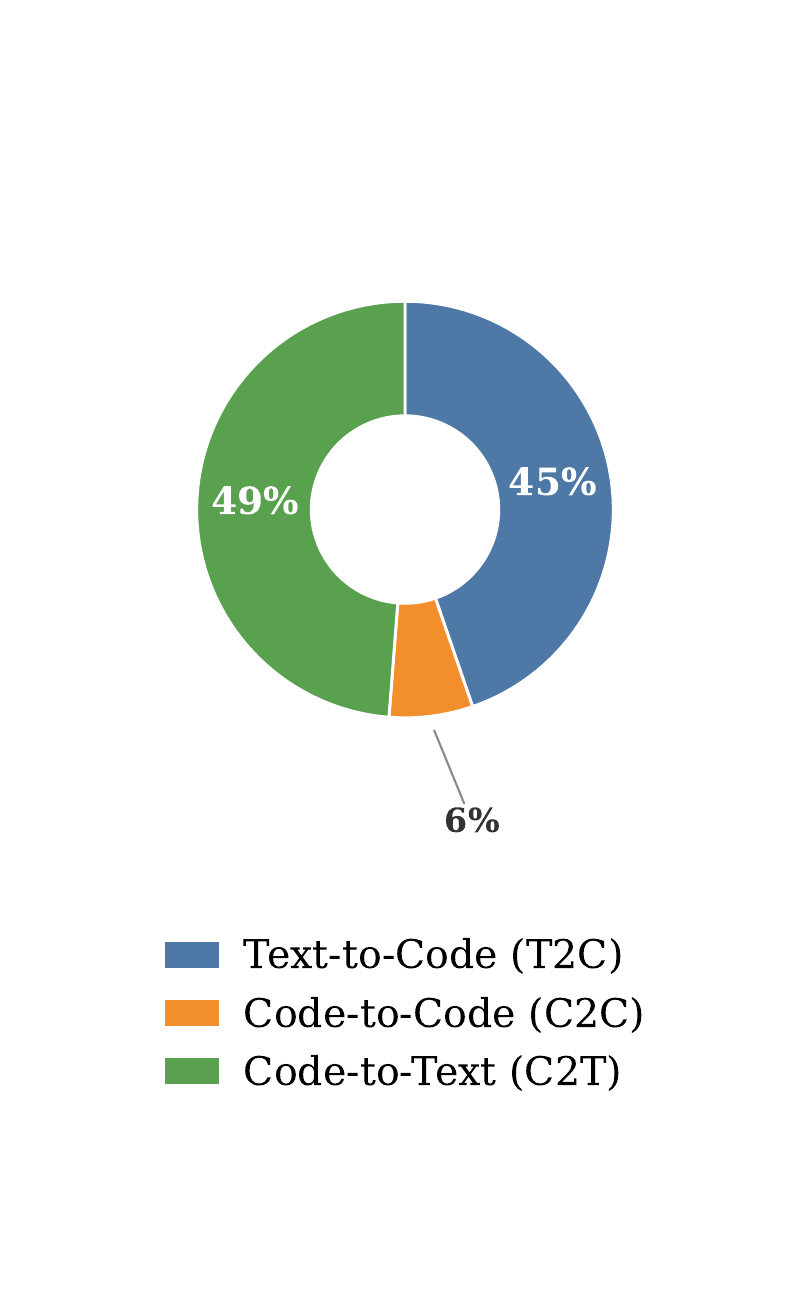}
    \caption{v202602 — by main task}
    \label{fig:task_overview_602_a}
  \end{subfigure}
  \hfill
  \begin{subfigure}[b]{0.32\linewidth}
    \centering
    \includegraphics[width=\linewidth]{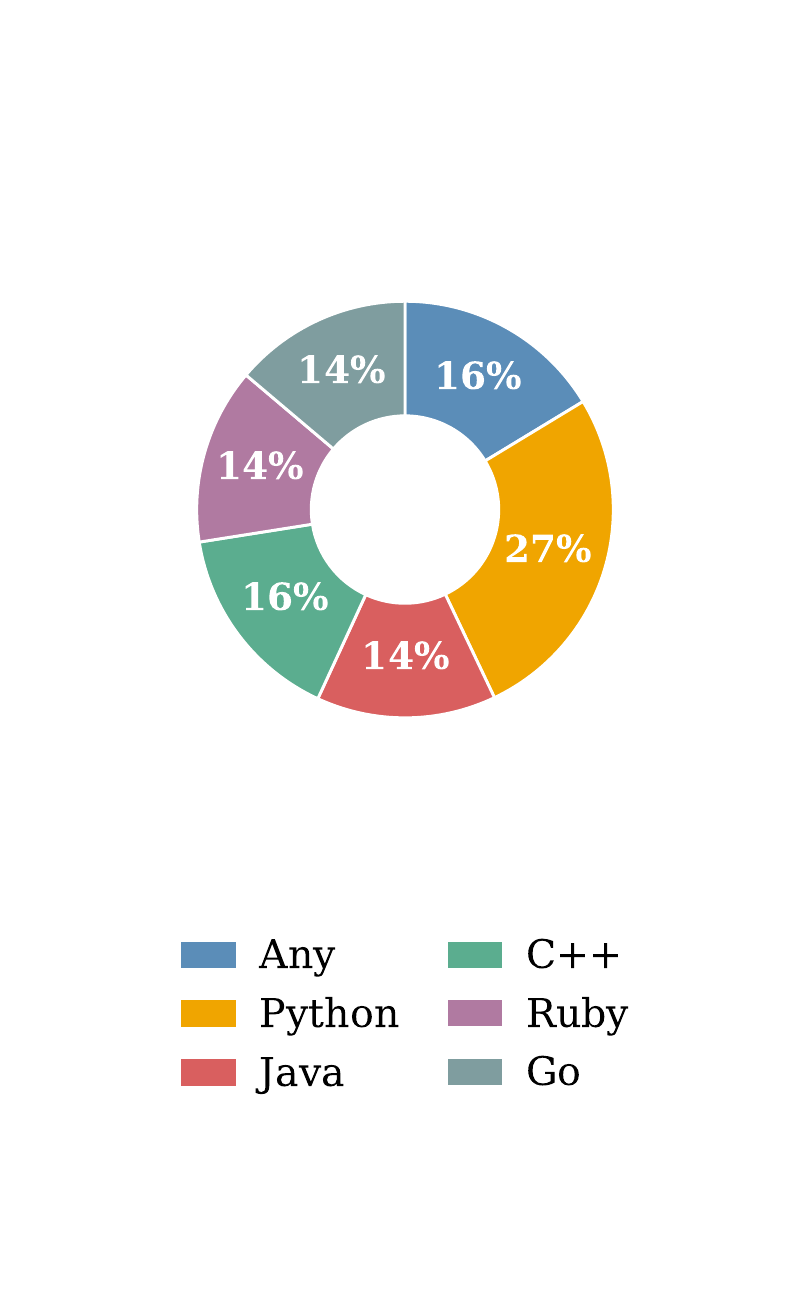}
    \caption{v202602 — by language}
    \label{fig:task_overview_602_b}
  \end{subfigure}
  \hfill
  \begin{subfigure}[b]{0.32\linewidth}
    \centering
    \includegraphics[width=\linewidth]{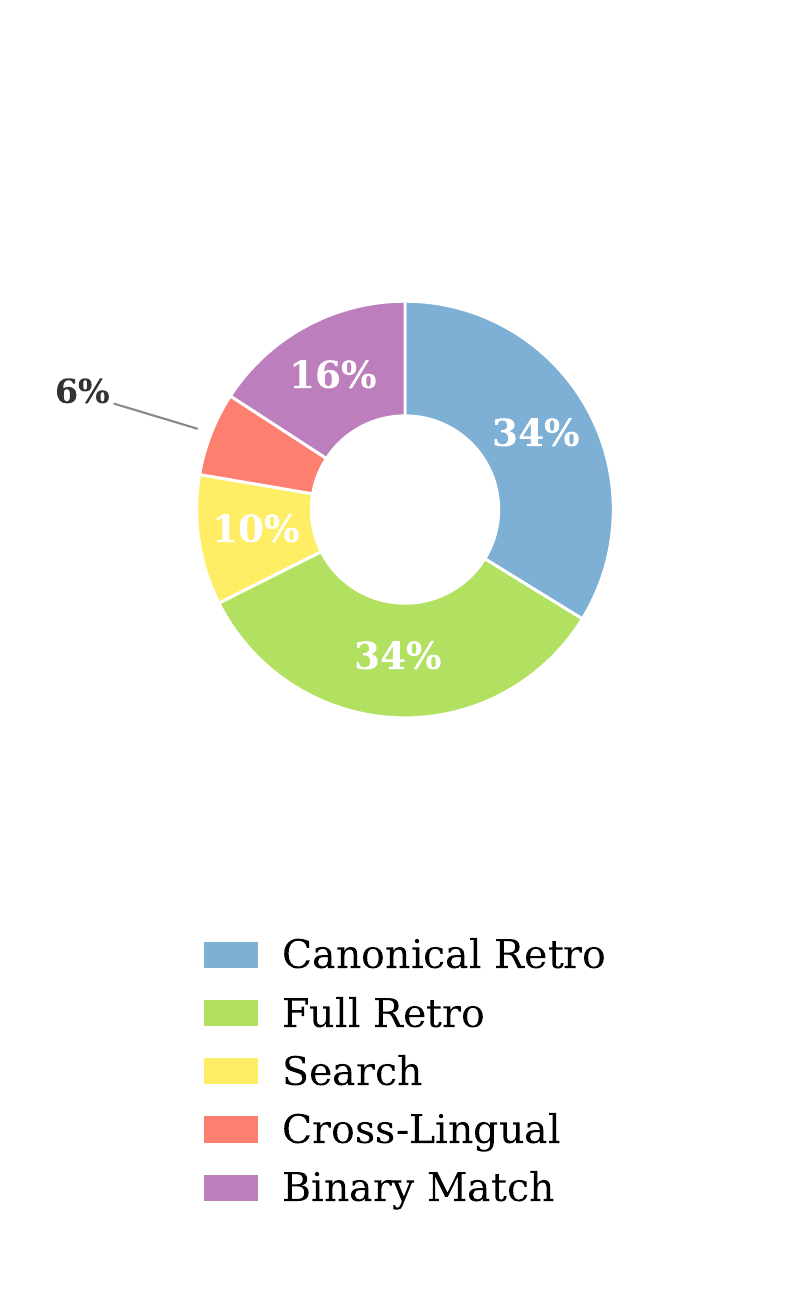}
    \caption{v202602 — by subtask}
    \label{fig:task_overview_602_c}
  \end{subfigure}

  \vspace{6pt}

  \begin{subfigure}[b]{0.32\linewidth}
    \centering
    \includegraphics[width=\linewidth]{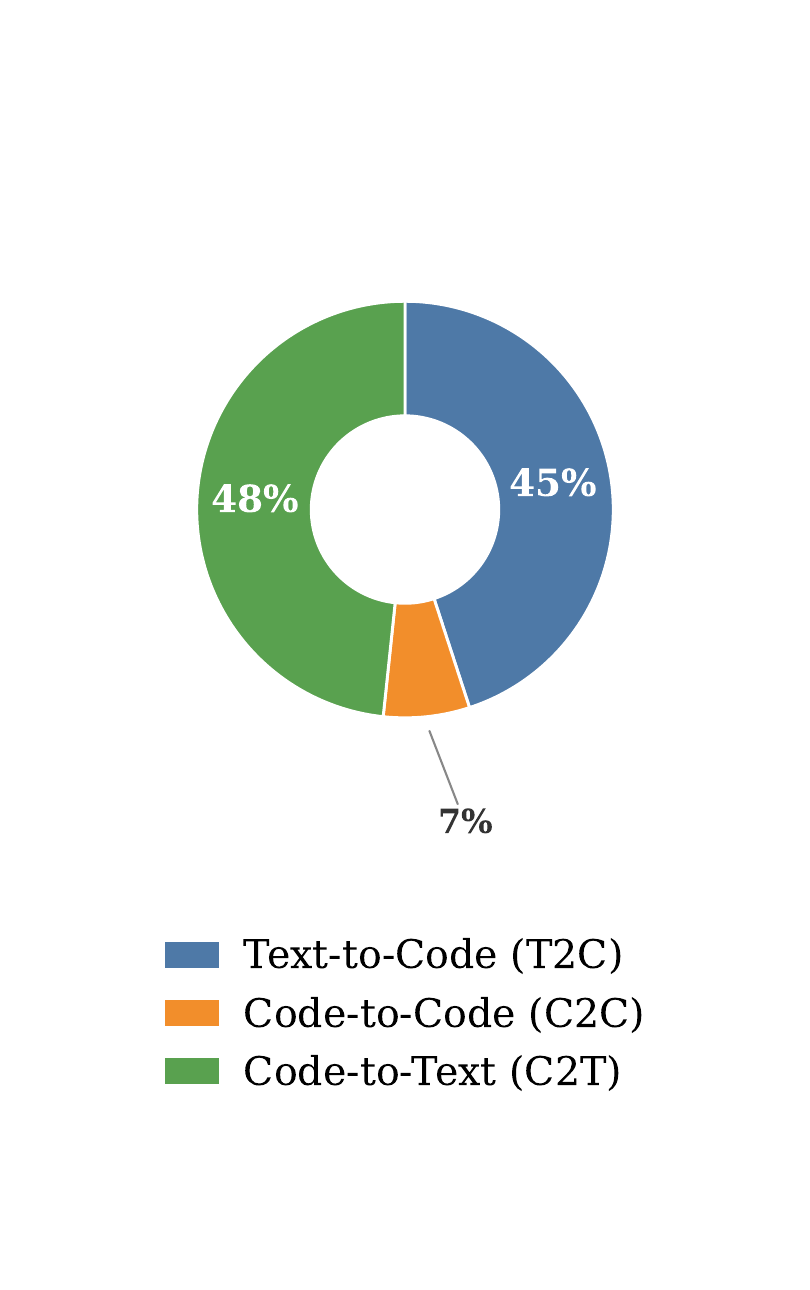}
    \caption{v202603 — by main task}
    \label{fig:task_overview_603_a}
  \end{subfigure}
  \hfill
  \begin{subfigure}[b]{0.32\linewidth}
    \centering
    \includegraphics[width=\linewidth]{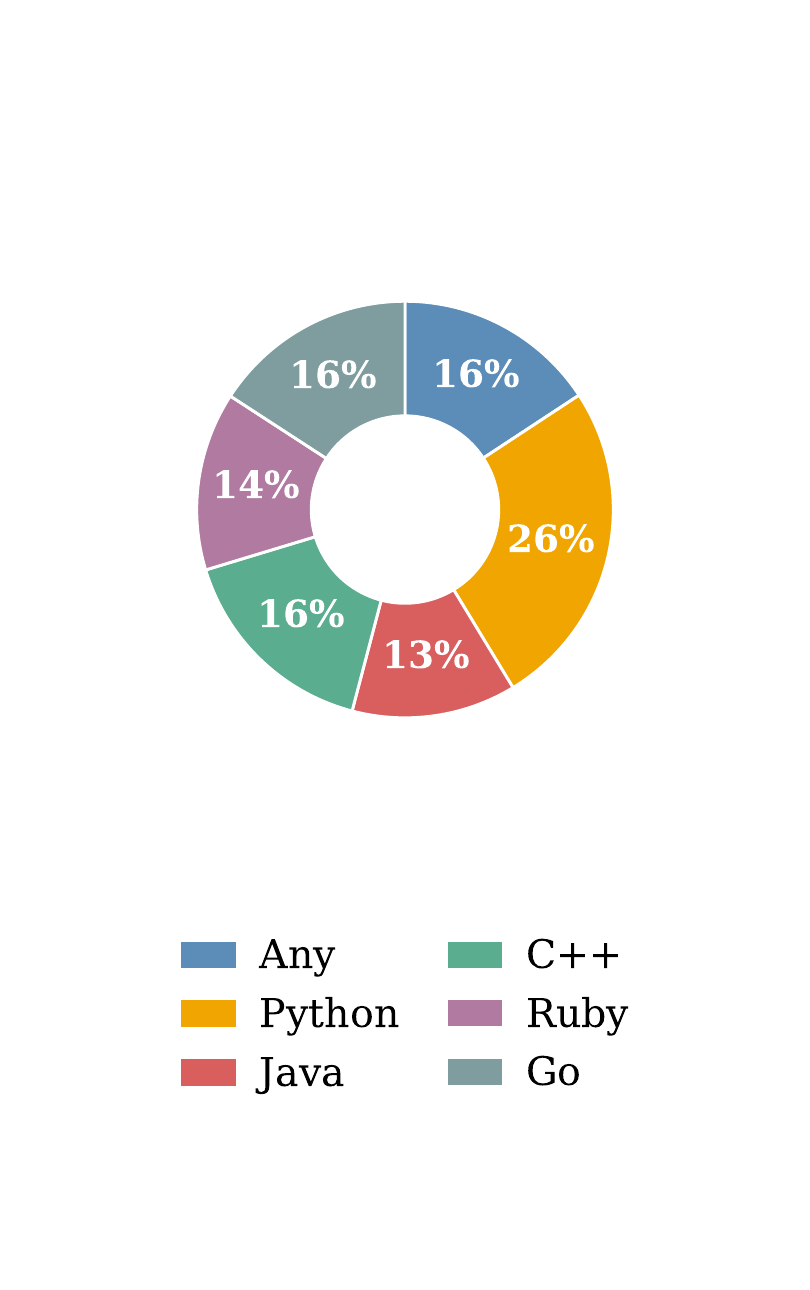}
    \caption{v202603 — by language}
    \label{fig:task_overview_603_b}
  \end{subfigure}
  \hfill
  \begin{subfigure}[b]{0.32\linewidth}
    \centering
    \includegraphics[width=\linewidth]{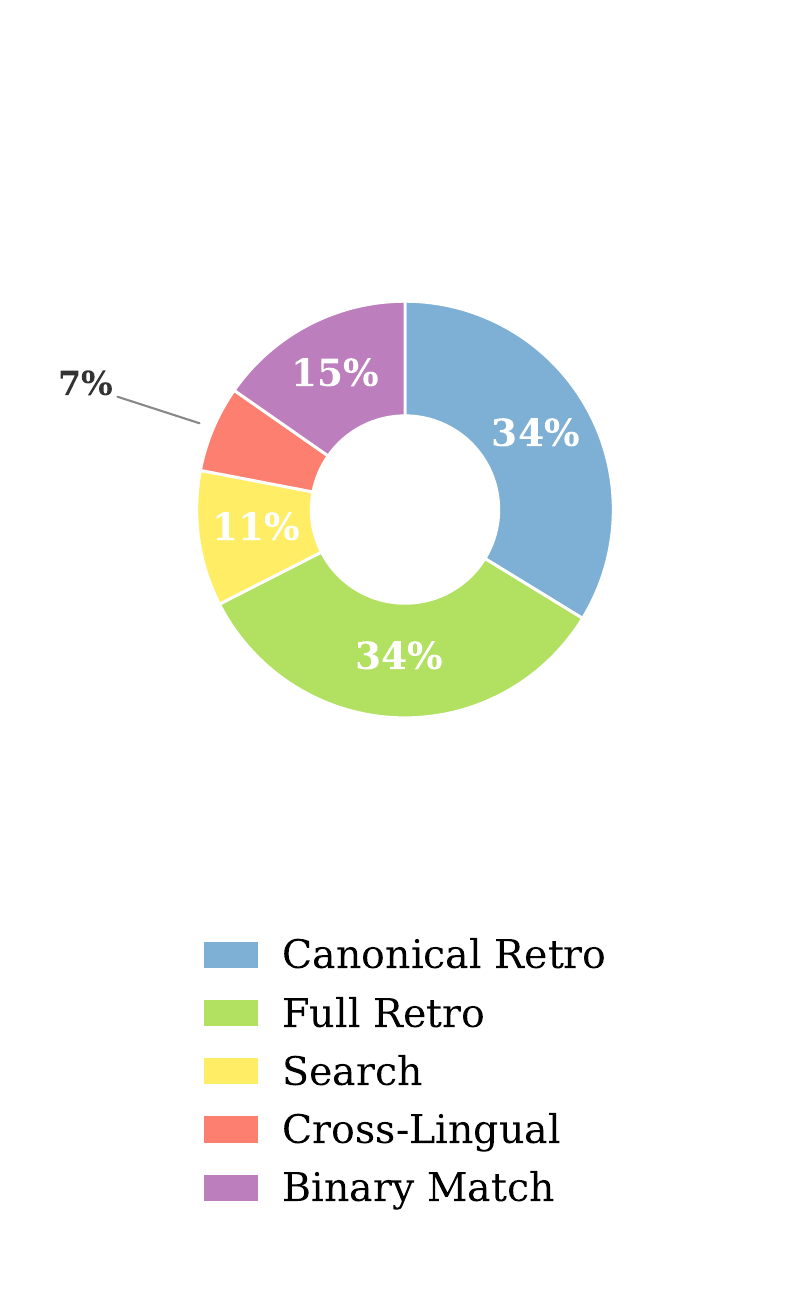}
    \caption{v202603 — by subtask}
    \label{fig:task_overview_603_c}
  \end{subfigure}

  \caption{Query distribution per \textsc{CoREB} release: \texttt{v202602} (2{,}604 queries,
  top row) and \texttt{v202603} (2{,}483 queries, bottom row). The two releases exhibit
  nearly identical proportions; see \cref{tab:query_dist_summary} for exact counts.}
  \label{fig:task_overview_per_release}
\end{figure*}

\begin{table}[t]
  \centering
  \small
  \renewcommand{\arraystretch}{1.2}
  \setlength{\tabcolsep}{6pt}
  \begin{sc}
  \begin{tabular}{l l r r r}
    \toprule
    \textbf{Corpus} & \textbf{Type} & \textbf{v202602} & \textbf{v202603} & \textbf{Aggregated} \\
    \midrule
    \multirow{6}{*}{Code Corpus}
      & Python   & 334 & 349 &   683 \\
    & Java       & 334 & 348 &   682 \\
    & C++        & 334 & 348 &   682 \\
    & Go         & 334 & 349 &   683 \\
    & Ruby       & 334 & 350 &   684 \\
    \cmidrule{2-5}
    & \textbf{Total} & \textbf{1{,}670} & \textbf{1{,}744} & \textbf{3{,}414} \\
    \midrule
    \multirow{4}{*}{Text Corpus}
      & Original descriptions  & 167 & 175 &   342 \\
    & Abbreviated descriptions & 167 & 175 &   342 \\
    & Hard negatives           & 668 & 700 & 1{,}368 \\
    \cmidrule{2-5}
    & \textbf{Total} & \textbf{835} & \textbf{875} & \textbf{1{,}710} \\
    \bottomrule
  \end{tabular}
  \end{sc}
  \caption{Composition of code and text corpora in \textsc{CoREB} per release and aggregated across both releases. Each release regenerates candidates with its own code models and draws from a freshly sourced LiveCodeBench snapshot, so aggregated artifacts are sums of distinct records (no cross-release ID overlap). In \texttt{v202603}, 537 of 1{,}744 generated code candidates pass every test case (verified-correct); the remainder carry \texttt{passed\,=\,false} metadata (see \cref{app:solution_gen} for the per-model breakdown). Hard negatives are generated with four strategies (operation-type change, objective change, algorithmic-approach change, and problem-domain change); see \cref{app:noise_gen}.}
  \label{tab:corpus_composition}
\end{table}

\subsection{Dataset Format and Naming Convention}
\label{app:coreb_format}
\paragraph{Query data format.}
The triple \texttt{(main\_task, subtask\_type, language\_constraint)} is encoded in the \emph{query identifier} (not the subtask definition itself). Specifically, each query ID follows:
\begin{center}
\texttt{q\_\{main.task\}\_\{subtask.type\}\_\{language.constraint\}\_\{index\}},
\end{center}
where \texttt{main.task} $\in \{\texttt{t2c}, \texttt{c2c}, \texttt{c2t}\}$, \texttt{subtask.type} is the subtask-type label, \texttt{language.constraint} $\in \{\texttt{any}, \texttt{python}, \texttt{java}, \texttt{cpp}, \texttt{ruby}, \texttt{go}\}$, and \texttt{index} is a zero-padded unique number. For example, \texttt{q\_t2c\_full\_retro\_python\_0001} denotes a text-to-code query of type \texttt{full\_retro} with a Python target-language constraint. \Cref{lst:query_example} lists the full field schema; concrete per-task examples are shown in \cref{app:task_examples}.

\begin{lstlisting}[caption={JSON schema of a text-to-code query record. See \cref{app:task_examples} for full per-task examples.},label={lst:query_example}]
{
  "query_id":          "<string>",  // e.g. q_t2c_canonical_retro_any_0001
  "query":             "<string>",  // natural-language problem description
  "split":             "<string>",  // text2code | code2code | code2text
  "subtask":           "<string>",  // e.g. t2c_canonical_retro_any
  "query_type":        "<string>",  // language_agnostic | language_specific
  "sub_query_type":    "<string>",  // abbreviated | full_description | ...
  "language_constraint": "<string>",// none | python | java | cpp | ruby | go
  "meta": {
    "source_problem_id": "<string>" // originating LiveCodeBench problem ID
  }
}
\end{lstlisting}

\paragraph{Text corpus data format.}
The text corpus record, illustrated in \Cref{lst:text_corpus_example}, stores two aligned natural-language views of the same seed problem. The field \texttt{text} contains the full problem statement (used in text-to-code queries when \texttt{sub\_query\_type=full\_description}), while \texttt{abbreviated\_text} provides a condensed version for the canonical retro setting (used when \texttt{sub\_query\_type=abbreviated}). The attribute \texttt{text\_style} specifies the formatting template of the full statement (e.g., title plus description), and the lengths \texttt{text\_length} and \texttt{abbreviated\_text\_length} are tracked for analysis of length sensitivity. Finally, \texttt{meta.source\_problem\_id} links the entry back to the originating seed problem, enabling grouping and consistency checks across variants. 

\begin{itemize}[leftmargin=*, itemsep=0.2em]
  \item \texttt{text} provides the full problem statement (used by text-to-code queries with \texttt{sub\_query\_type=full\_description}).
  \item \texttt{abbreviated\_text} is a condensed version of the same problem (used by text-to-code queries with \texttt{sub\_query\_type=abbreviated}).
  \item \texttt{text\_style} indicates the formatting template of the full text (e.g., title + description).
  \item \texttt{text\_length} and \texttt{abbreviated\_text\_length} track query length for analysis and stratification.
  \item \texttt{meta.source\_problem\_id} links the text entry back to the original seed problem, enabling grouping and deduplication across variants.
\end{itemize}

\begin{lstlisting}[
  style=corebjson,
  caption={Example text corpus record for text-to-code: full description and its abbreviated form with metadata.},
  label={lst:text_corpus_example}
]
{
  "text_id": "text_v202601_00001",
  "text_style": "title_plus_description",
  "text": "Determine Maximum Removals from Playlist While Preserving Favorite Sequence\n"
          "You are given a playlist `playlist` of size n, a list `favoriteSequence` that is a subsequence of `playlist`, "
          "and a sorted integer array `removableIndices` with distinct values in [0, n-1].\n"
          "An operation removes a song at index idx such that idx is in `removableIndices` and `favoriteSequence` remains "
          "a subsequence after removal.\n"
          "Performing an operation does not change the indices of the other songs in `playlist`...\n",
  "text_length": 2076,
  "abbreviated_text": "Given a playlist `playlist` of size n, a subsequence `favoriteSequence`, and a sorted array "
                      "`removableIndices` of distinct indices in [0, n-1], find the maximum number of songs that can be "
                      "removed while ensuring `favoriteSequence` remains a subsequence...",
  "abbreviated_text_length": 562,
  "meta": {
    "source_problem_id": "lcb_3487"
  }
}
\end{lstlisting}

\paragraph{Code corpus data format.}
The code-corpus record shown in \Cref{lst:code_corpus_example} represents a single candidate implementation tied to a seed problem via
\texttt{meta.source\_problem\_id}. The raw program text is stored in \texttt{code}, while \texttt{language}
and \texttt{model} identify the programming language and the generator (or source) of the solution,
which enables code-to-code evaluation along the cross-language axis. The field
\texttt{code\_length} is recorded for length-based analyses. Finally, \texttt{meta} captures execution
results from the unit-test harness (\texttt{passed}, \texttt{pass\_rate}, \texttt{test\_passed}/\texttt{test\_total}).
The corpus retains all 1,744 generated candidates, both correct and incorrect, so that the \texttt{passed} field can be used for task-specific relevance labeling: code-to-code tasks restrict positives to verified-correct solutions, while text-to-code tasks treat all solutions for the queried problem as relevant (see \cref{sec:benchmark}).

\begin{lstlisting}[
  style=corebjson,
   caption={Example code-corpus record: a candidate solution with language/model metadata and execution outcomes.  This entry has \texttt{passed\,=\,false}; the corpus retains such entries alongside verified-correct ones (see text).},
  label={lst:code_corpus_example}
]
{
  "code_id": "code_v202601_00001",
  "code": "using namespace std;\n\nclass Solution {\npublic:\n    int maxRemovals(string playlist, string favoriteSequence, vector<int>& removableIndices) {\n        int n = playlist.length();\n        int m = favoriteSequence.length();\n        \n        auto canRemove = [&](int k) -> bool {\n            set<int> removed;\n            for (int i = 0; i < k; i++) {\n                removed.insert(removableIndices[i]);\n            }\n            \n            int j = 0;\n            for (int i = 0; i < n && j < m; i++) {\n                if (removed.count(i)) continue;\n                if (playlist[i] == favoriteSequence[j]) {\n                    j++;\n                }\n            }\n            return j == m;\n        };\n        \n        int left = 0, right = removableIndices.size();\n        int result = 0;\n        \n        while (left <= right) {\n            int mid = (left + right) / 2;\n            if (canRemove(mid)) {\n                result = mid;\n                left = mid + 1;\n            } else {\n                right = mid - 1;\n            }\n        }\n        \n        return result;\n    }\n};\n\nint main() {\n    string playlist, favoriteSequence;\n    cin >> playlist >> favoriteSequence;\n    \n    int k;\n    cin >> k;\n    vector<int> removableIndices(k);\n    for (int i = 0; i < k; i++) {\n        cin >> removableIndices[i];\n    }\n    \n    Solution sol;\n    cout << sol.maxRemovals(playlist, favoriteSequence, removableIndices) << endl;\n    \n    return 0;\n}\n\nmain();",
  "language": "cpp",
  "model": "claude-sonnet-4-5",
  "code_length": 1460,
  "meta": {
    "source_problem_id": "lcb_3487",
    "solution_key": "claude-sonnet-4-5_cpp",
    "passed": false,
    "pass_rate": 0.0,
    "test_passed": 0,
    "test_total": 1
  }
}
\end{lstlisting}

\subsection{Task Instance Examples}
\label{app:task_examples}

The following listings show representative query instances and their relevance judgments (qrels) for each of the three main tasks.
All examples are drawn directly from the CoREB dataset.
Code snippets are truncated for readability; the full texts are available in the released data files.

\paragraph{text-to-code.}
text-to-code queries are natural-language problem descriptions targeting a code corpus.
\Cref{lst:t2c_canonical_ex} shows an abbreviated (canonical retro) query with no language constraint.
\Cref{lst:t2c_search_ex} shows a short LLM-generated developer-style search query restricted to Python.

\begin{lstlisting}[
  style=corebjson,
  caption={text-to-code example: abbreviated (canonical retro) query with language-agnostic constraint and its qrel.},
  label={lst:t2c_canonical_ex}]
// --- Query ---
{
  "query_id": "q_t2c_canonical_retro_any_0001",
  "query": "Given m arenas and t days, find maximum points a player can
    accumulate. Each day the player can stay (earn restBonus[i][current])
    or switch arena (earn switchBonus[current][dest]). Player may start
    at any arena. Constraints: 1<=m,t<=200.",
  "split": "text2code",
  "subtask": "t2c_canonical_retro_any",
  "query_type": "language_agnostic",
  "sub_query_type": "abbreviated",
  "language_constraint": "none",
  "meta": { "source_problem_id": "lcb_3587" }
}

// --- Qrel ---
{ "query_id": "q_t2c_canonical_retro_any_0001",
  "doc_id":   "code_v202601_00028",
  "relevance": 1 }
\end{lstlisting}

\begin{lstlisting}[
  style=corebjson,
  caption={text-to-code example: LLM-generated developer-style search query with Python language constraint and its qrel.},
  label={lst:t2c_search_ex}]
// --- Query ---
{
  "query_id": "q_t2c_search_python_0787",
  "query": "Find the Kth character after repeatedly inverting case
    and concatenating a string 10^100 times",
  "split": "text2code",
  "subtask": "t2c_search_python",
  "query_type": "language_specific",
  "sub_query_type": "language_agnostic",
  "language_constraint": "python",
  "meta": { "source_problem_id": "lcb_abc380_d" }
}

// --- Qrel ---
{ "query_id": "q_t2c_search_python_0787",
  "doc_id":   "code_v202601_00...",
  "relevance": 1 }
\end{lstlisting}

\paragraph{code-to-code.}
code-to-code queries are code snippets; the goal is to retrieve semantically equivalent implementations.
The current release contains only the cross-language subtask.
\Cref{lst:c2c_cross_ex} shows a cross-language query (Java anchor by Claude~Sonnet~4.5, Python target).

\begin{lstlisting}[
  style=corebjson,
  caption={code-to-code example: cross-language query (Java anchor by Claude~Sonnet~4.5, Python target) and its qrel.},
  label={lst:c2c_cross_ex}]
// --- Query ---
{
  "query_id": "q_c2c_cross_lang_0002",
  "query": "import java.util.*;\n\npublic class Solution {\n    public
    static void main(String[] args) {\n        Scanner sc = new
    Scanner(System.in);\n        ...  (truncated)",
  "split": "code2code",
  "subtask": "c2c_cross_lang",
  "anchor_language": "java",
  "anchor_model": "claude-sonnet-4-5",
  "meta": {
    "source_problem_id": "lcb_3616",
    "anchor_code_id": "code_v202601_00173",
    "target_languages": ["python"],
    "num_target_solutions": 1
  }
}

// --- Qrel ---
{ "query_id": "q_c2c_cross_lang_0002",
  "doc_id":   "code_v202601_00174",
  "relevance": 1 }
\end{lstlisting}

\paragraph{code-to-text.}
code-to-text queries are code snippets; the retrieval target is a natural-language problem description.
\Cref{lst:c2t_retro_ex} shows a full-retro retrieval query. \Cref{lst:c2t_match_ex} shows a pair-match
instance, which provides a pre-paired (code, text) with a binary label; evaluation still ranks the full text corpus.

\begin{lstlisting}[
  style=corebjson,
  caption={code-to-text example: full-retro retrieval query (Python, language-agnostic) and its qrel.},
  label={lst:c2t_retro_ex}]
// --- Query ---
{
  "query_id": "q_c2t_full_retro_any_0001",
  "query": "from typing import List\n\nclass Solution:\n    def
    maxPoints(self, m: int, t: int, restBonus: List[List[int]],
    switchBonus: List[List[int]]) -> int:\n        dp = [[0]*m for
    _ in range(t+1)]\n        ...  (truncated)",
  "split": "code2text",
  "subtask": "c2t_full_retro_any",
  "anchor_language": "python",
  "anchor_model": "claude-sonnet-4-5",
  "meta": {
    "source_problem_id": "lcb_3587",
    "anchor_code_id": "code_v202601_00028",
    "task_type": "description_retrieval",
    "text_type": "full_description"
  }
}

// --- Qrel ---
{ "query_id": "q_c2t_full_retro_any_0001",
  "doc_id":   "text_v202601_00001",
  "relevance": 1 }
\end{lstlisting}

\begin{lstlisting}[
  style=corebjson,
  caption={code-to-text example: pair-match instance. The code and a specific text entry are pre-paired; label=1 means they correspond (positive pair). Evaluation still follows the standard retrieval protocol: the code is used as a query against the full text corpus.},
  label={lst:c2t_match_ex}]
{
  "query_id": "q_c2t_match_0641",
  "code": "from typing import List\n\nclass Solution:\n    def
    maxPoints(self, m: int, t: int, ...) -> int:\n        ...
    (truncated)",
  "text_id": "text_v202601_00001",
  "split": "code2text",
  "subtask": "c2t_match",
  "label": 1,
  "text_type": "full",
  "anchor_language": "python",
  "anchor_model": "claude-sonnet-4-5",
  "meta": {
    "source_problem_id": "lcb_3587",
    "anchor_code_id": "code_v202601_00028",
    "pair_type": "positive"
  }
}
\end{lstlisting}

\subsection{Evaluation Protocol and Metrics}
\label{app:coreb_metrics}

All subtasks, including \texttt{c2t\_match}, are evaluated with a single, uniform ranked-retrieval protocol.
Given a query, every model encodes it and ranks all items in the corresponding corpus by cosine similarity; the resulting ranked list is scored with
\textbf{nDCG@k} ($k\!=\!1,3,5,10$),
\textbf{MAP@k} ($k\!=\!10,100$),
\textbf{Recall@k} ($k\!=\!1,3,5,10,100$),
\textbf{MRR@k} ($k\!=\!10$), and
\textbf{Precision@k} ($k\!=\!1,3,5,10$).
Negative-pair instances in \texttt{c2t\_match} have only hard-negative qrels (rel=1) and no true positives (rel=2); under \texttt{relevance\_level=2} they contribute a score of zero, acting as a fixed penalty that rewards models only for the positive-pair and retro subtasks.

\paragraph{Graded relevance and evaluation strictness.}
Starting from \textsc{CoREB} v2, qrels use a three-level graded relevance scheme: \texttt{relevance=2} (true positive), \texttt{relevance=1} (hard negative), and absent (easy negative / unjudged).
All metrics are computed with \texttt{relevance\_level=2}: standard binary metrics (Recall, MAP, Precision, MRR) treat only rel=2 items as relevant.
For nDCG, hard negatives are zeroed to rel=0 before scoring so that they contribute zero gain regardless of rank, yet they still penalize by occupying positions that true positives could occupy.
This design is strictly harder than benchmarks using binary qrels (e.g., BEIR, CoIR, MTEB) where all explicitly judged items count as relevant: under those schemes, a model retrieving a failed code solution or a noise description would receive credit; under our graded scheme, it would not.
Evaluation is performed via \texttt{pytrec\_eval} with the filtered qrels passed as-is to the relevance evaluator.

\paragraph{code-to-code anchor exclusion.}
For the code-to-code subtask, the anchor code item (i.e., the exact code snippet used as the query) is always present in the shared retrieval corpus.
Because every model assigns near-perfect similarity to the anchor, it is trivially retrieved at rank~1 without providing any signal about cross-language retrieval ability.
We therefore remove the anchor from each query's ranked list before computing any metric; the reported nDCG@$k$, Recall@$k$, and MRR@$k$ values for code-to-code thus reflect retrieval quality over the remaining corpus positions (see \cref{app:anchor_exclusion} for full details).

\subsection{Experimental Validation: Effect of Annotation}
\label{app:annotation_quality}

To assess how our annotation pipeline affects evaluation reliability, we run a controlled $2{\times}2$ comparison: two frontier models (Gemini~3~Flash and Claude~Sonnet~4.5) evaluated on two LiveCodeBench releases (\texttt{release\_v5} / \texttt{v202602} and \texttt{release\_v6} / \texttt{v202603}).
The two releases cover non-overlapping contest windows: \texttt{v202602} spans Sep~2024–Jan~2025, while \texttt{v202603} spans Jan~2025–Apr~2025.
Both models share a nominal knowledge cutoff of \textbf{January 2025}: Gemini~3~Flash (released Dec~17, 2025; knowledge cutoff Jan~2025~\citep{gemini3flash2025}) and Claude~Sonnet~4.5 (released Sep~29, 2025; reliable knowledge cutoff Jan~2025, training data cutoff Jul~2025~\citep{claude2025sonnet45}).
For each problem, we prompt each model using \cref{lst:code_gen_prompt} to generate a single solution in each target language and score it against the corresponding test suite.
We report Pass@1, i.e., the fraction of problems for which the generated solution passes all test cases.

\paragraph{Results.}
\Cref{tab:annotation_pass1} reports Pass@1 on original and annotated problems across all four model--release combinations.

Gemini~3~Flash shows a consistent drop in \emph{both} releases: $-8.6$ points on \texttt{v202602} ($43\%\to34\%$) and $-6.5$ points on \texttt{v202603} ($38\%\to31\%$), across all five languages without exception.
Because \texttt{v202602} problems fall squarely within Gemini's training window (before Jan~2025), leakage is strongest there; the smaller but still consistent drop on \texttt{v202603} (starting Jan~4, 2025) suggests Gemini's training data includes problems right up to its nominal cutoff.

Claude~Sonnet~4.5 reveals a striking \emph{asymmetry}: Pass@1 drops by $12.3$ points on \texttt{v202602} ($35\%\to22\%$) — a larger drop than Gemini on the same release — yet shows virtually no change on \texttt{v202603} ($29\%\to31\%$, $+2.0$ points).
This dissociation is directly explained by Claude's \emph{reliable} knowledge cutoff of January~2025: the \texttt{v202602} contest problems (ending Jan~4, 2025) lie entirely within Claude's reliably memorized window, whereas the \texttt{v202603} problems (starting Jan~4, 2025) fall beyond it, yielding near-zero leakage.

Taken together, the two models yield complementary evidence.
Gemini leaks on both releases; Claude leaks strongly on one but not the other.
The pattern is model-dependent and release-dependent, and unpredictable without a controlled test: a benchmark builder cannot know a priori which problems any given model has memorized.
This is precisely why counterfactual annotation is necessary.
Using original problem statements in an embedding benchmark would introduce an uncontrolled performance bias; our rewriting pipeline eliminates it across both models and both releases.

Two secondary factors may also contribute modestly: (i) reformulation can reduce underspecification present in original statements, and (ii) test-suite validation during annotation eliminates solutions that passed only because of weak original tests.

\begin{table}[h]
  \centering
  \small
  \renewcommand{\arraystretch}{1.15}
  \setlength{\tabcolsep}{6pt}
  \begin{sc}
  \begin{tabular}{l rrr}
    \toprule
    Language & Orig & Ann & $\Delta$ \\
    \midrule
    \multicolumn{4}{l}{\textit{Gemini~3~Flash  v202602 (Sep 2024–Jan 2025)}} \\
    \quad Python  & .491 & .377 & \textminus.114 \\
    \quad Java    & .293 & .251 & \textminus.042 \\
    \quad C++     & .425 & .383 & \textminus.042 \\
    \quad Go      & .407 & .305 & \textminus.102 \\
    \quad Ruby    & .515 & .383 & \textminus.132 \\
    \quad Overall & .426 & .340 & \textminus.086 \\
    \midrule
    \multicolumn{4}{l}{\textit{Gemini~3~Flash  v202603 (Jan 2025–Apr 2025)}} \\
    \quad Python  & .411 & .337 & \textminus.074 \\
    \quad Java    & .274 & .217 & \textminus.057 \\
    \quad C++     & .360 & .349 & \textminus.011 \\
    \quad Go      & .354 & .314 & \textminus.040 \\
    \quad Ruby    & .480 & .337 & \textminus.143 \\
    \quad Overall & .376 & .311 & \textminus.065 \\
    \midrule
    \multicolumn{4}{l}{\textit{Claude~Sonnet~4.5  v202602 (Sep 2024–Jan 2025)}} \\
    \quad Python  & .395 & .246 & \textminus.150 \\
    \quad Java    & .281 & .180 & \textminus.102 \\
    \quad C++     & .347 & .246 & \textminus.102 \\
    \quad Go      & .371 & .240 & \textminus.132 \\
    \quad Ruby    & .341 & .210 & \textminus.132 \\
    \quad Overall & .347 & .224 & \textminus.123 \\
    \midrule
    \multicolumn{4}{l}{\textit{Claude~Sonnet~4.5  v202603 (Jan 2025–Apr 2025)}} \\
    \quad Python  & .345 & .326 & \textminus.019 \\
    \quad Java    & .241 & .257 & +.016 \\
    \quad C++     & .259 & .286 & +.027 \\
    \quad Go      & .276 & .326 & +.050 \\
    \quad Ruby    & .306 & .320 & +.014 \\
    \quad Overall & .285 & .303 & +.017 \\
    \bottomrule
  \end{tabular}
  \end{sc}
  \caption{Pass@1 on original vs.\ annotated problems across a $2{\times}2$ design (two models $\times$ two releases). $\Delta = \text{Ann} - \text{Orig}$. Gemini~3~Flash drops consistently in both releases; Claude~Sonnet~4.5 drops sharply on \texttt{v202602} (older problems, likely in training data) but shows no drop on \texttt{v202603} (newer problems, likely past training cutoff). The model-dependent, release-dependent pattern demonstrates that data leakage is unpredictable without controlled annotation.}
  \label{tab:annotation_pass1}
\end{table}

\subsection{Code Solution Generation}
\label{app:solution_gen}

The code corpus underlying \textsc{CoREB} is regenerated in each release by two frontier LLMs: Gemini~3~Flash and  Claude~Sonnet~4.5
For each annotated problem, both models are prompted to produce a single solution in each of the five target languages (Python, Java, C++, Go, Ruby). In \texttt{v202603}, this yields 1{,}744 candidate solutions (the nominal $175 \times 5 \times 2 = 1{,}750$ minus 6 missing model--language combinations); in \texttt{v202602}, the corresponding count is 1{,}670 (the nominal $167 \times 5 \times 2 = 1{,}670$, all combinations present).
Every candidate is executed against the full test suite; the outcome (\texttt{passed}, \texttt{pass\_rate}) is recorded as metadata.
All candidates are retained so that each problem has a (nearly) complete set of solutions across languages and models. The verified-correct counts (passing every test) are 528 of 1{,}670 in \texttt{v202602} (31.6\%) and 537 of 1{,}744 in \texttt{v202603} (30.8\%), for a total of 1{,}065 verified-correct solutions aggregated across releases.
Relevance labels are task-specific: code-to-code tasks restrict positives to verified-correct solutions, while text-to-code tasks treat all solutions for the queried problem as relevant.

The two models exhibit complementary strengths across languages (\cref{tab:annotation_pass1}), ensuring that few problems are left without any correct solution and that every language is represented in the verified-correct subset used for code-to-code relevance labeling.

\section{Experiment Details}
\label[appendix]{app:exp}

\subsection{Implementation Details}\label{app:imp}
For the analysis and data processing framework, we use the code from  the  public Github  repository at {\small \url{https://github.com/wuji3/Doraemon}}~\citep{du2025visual} with GNU General Public License.

For the evaluation framework, we use the code from  the  public Github  repository at {\small \url{https://github.com/beir-cellar/beir}}~\citep{thakur2021beir} and {\small \url{https://github.com/SerendipityOneInc/look-bench}} ~\citep{gao2026lookbench,xue2026quitobench} with Apache License and   {\small \url{https://github.com/eosphoros-ai/DB-GPT-Hub}}~\citep{xue2023dbgpt,xue2024demonstration,zhou2024dbgpthub} with MIT License.

\subsection{Inference and Evaluation Details}\label{app:training_details}

All embedding models are evaluated in inference-only mode; no fine-tuning or adaptation is performed on CoREB data.
For each model, we encode all corpus documents and queries with the model's default pooling strategy (e.g., CLS token, mean pooling, or instruction-prefixed mean pooling as specified by the model authors).
Similarity scores are computed via dot product or cosine similarity, depending on each model's recommended configuration.
Retrieval rankings are produced by scoring every (query, corpus entry) pair; dense nearest-neighbor search is used for efficiency where corpus size permits.
All experiments are run on a single NVIDIA RTX 4090 GPU with 24\,GB VRAM\@.
We report nDCG@10 and Recall@10 as primary metrics; full metric tables including MRR@10, Recall@$k$ ($k\!\in\!\{1,3,5,10,100\}$), and nDCG@$k$ ($k\!\in\!\{1,3,5,10\}$) are available via the project repository.

\subsection{Fine-Tuned \textsc{CoREB-Reranker} Training Protocol}
\label{app:reranker_training}

This appendix documents the full fine-tuning protocol for the \textsc{CoREB-Reranker} (4B) model introduced in \cref{sec:reranker}. The checkpoint will be released alongside the benchmark.

\paragraph{Architecture and base models.}

We initialise from the publicly released Qwen3-Reranker-4B model~\citep{qwen3embedding2025}. Unlike conventional cross-encoders with a scalar classification head, Qwen3-Reranker is used as a causal-LM reranker: for each (query, document) pair, the model is prompted to answer whether the document satisfies the query, and relevance is represented by the probability assigned to the verbal answer \texttt{yes}. We retain the released tokenizer, chat template, instruction format, and architectural configuration. Fine-tuning is parameter-efficient: we attach LoRA adapters to all linear modules with rank and alpha to 16 and dropout 0.05, while keeping the base model weights frozen.

\paragraph{Training data construction.}
Training uses a mixed reranker corpus consisting of \texttt{CoREB\_v202602}, CodeSearchNet~(including code-to-code, code-to-text, and text-to-code retrieval~(\citealp{2019CodeSearchNet,2025CoIR,2021CodeXGLUE}), \texttt{APPS}~\citep{2021APPS}, CosQA~\citep{2021CosQA}, and single-turn and multi-turn CodeFeedback~\citep{2024OpenCodeInterpreter}. Each JSONL record is normalized into independent binary reranking examples of the form
  \[
    (\text{instruction}, q, d, y), \qquad y \in \{\texttt{yes}, \texttt{no}\}.
  \]
  For each source record, positive documents are duplicated twice, and we sample one easy negative and one hard negative when available. Positive examples are labeled \texttt{yes}, while both easy and hard negatives are labeled \texttt{no}. The resulting examples are shuffled across tasks before training.
  We reserve a small held-out split for validation.
  
\paragraph{Prompt format and loss function.}
  We fine-tune Qwen3-Reranker in its generative yes/no reranking format. Each training example is serialized with the following system prompt:
  \begin{quote}\small
  \texttt{Judge whether the Document meets the requirements based on the Query and the Instruct provided. Note that the answer can only be "yes" or "no"}
  \end{quote}
  The user message is formatted as
  \begin{quote}\small
  \texttt{<Instruct>: \{instruction\}}\\
  \texttt{<Query>: \{query\}}\\
  \texttt{<Document>: \{document\}}
  \end{quote}
  where the instruction is task-specific. For text-to-code (T2C), we use:
  \begin{quote}\small
  \texttt{Given a natural language programming task, retrieve code that correctly solves or implements the task.}
  \end{quote}
  For code-to-code (C2C), we use:
  \begin{quote}\small
  \texttt{Given a code snippet, retrieve code that is semantically equivalent or solves the same task.}
  \end{quote}
  For code-to-text (C2T), we use:
  \begin{quote}\small
  \texttt{Given a code snippet, retrieve the natural language description or problem statement that best matches the code.}
  \end{quote}
  The assistant response is the verbal label $y \in \{\texttt{yes}, \texttt{no}\}$, with positives labeled \texttt{yes} and both easy and hard negatives labeled \texttt{no}. Training optimizes the causal language-modeling cross-entropy only over the assistant answer region:
  \[
    \mathcal{L}
    =
    - \frac{1}{|\mathcal{M}|}
    \sum_{t \in \mathcal{M}}
    \log p_{\theta}(x_t \mid x_{<t}),
  \]
  where $\mathcal{M}$ masks out all system and user prompt tokens and keeps only the answer tokens. Thus, hard negatives and easy negatives are not placed in a shared listwise denominator; they contribute as independent \texttt{no}-labeled pairwise training examples.

\paragraph{Optimization.}
We fine-tune for 2 epochs with AdamW ($\beta_1{=}0.9$, $\beta_2{=}0.999$, $\epsilon{=}10^{-8}$, weight decay $0.01$), a peak learning rate of $1{\times} 10^{-5}$, linear warm-up for 100 configured steps, and a linear learning-rate decay schedule. Training uses \texttt{bf16}, FlashAttention-2, gradient checkpointing, DeepSpeed ZeRO-2 through \texttt{accelerate}, and a per-device batch size of 12 with gradient accumulation 1. The maximum sequence length is 4{,}096 tokens, with right-side truncation and left padding. We train the reranker on 8 NVIDIA A100 GPUs.
\paragraph{Checkpoint merging.}
The released \textsc{CoREB-Reranker} checkpoint is the \emph{uniform model soup}~\citep{wortsman2022modelsoups} of two independently fine-tuned LoRA variants. Both variants share the same base model, LoRA configuration, optimizer, and training corpus, and differ only in random seed and data shuffle order. The soup is obtained by averaging their LoRA adapter weights with equal coefficients before merging into the base model. We selected the two variants by validation nDCG@10 on a held-out split prior to averaging; the soup outperforms either individual checkpoint on the mean of the three test tasks while matching their per-task best on T2C and C2T.

\paragraph{Evaluation.}
At test time, we (i) run C2LLM-7B retrieval on the \texttt{v202603} corpus to obtain the top-128 candidates per query, (ii) score each (query, candidate) pair with the trained reranker, and (iii) re-sort by the reranker score before computing nDCG@10 and Recall@10 against the graded qrels (\cref{app:metrics}). The reranker never sees \texttt{v202603} problems, queries, or qrels during training, so the evaluation is fully out-of-sample at the problem level.

\subsection{Evaluation Metrics}
\label[appendix]{app:metrics}
\textbf{Recall@k.} Recall is a commonly used metric in retrieval tasks, measuring the proportion of relevant items successfully retrieved within the top-$k$ results. Formally, given a query with $R$ relevant items in the dataset, Recall@k is defined as:
\begin{equation}
\mathrm{Recall@}k = \frac{1}{R} \sum_{i=1}^{k} rel(i),
\label{eq:recall}
\end{equation}
where $rel(i) \in \{0,1\}$ indicates whether the $i$-th retrieved item is relevant. Since Recall@k evaluates coverage rather than ordering, it ranges between 0 and 1 and increases monotonically with $k$.

\medskip
\textbf{nDCG@k.} Normalized Discounted Cumulative Gain (nDCG) is a weighted ranking metric that evaluates the quality of the ordering of retrieved items. The discounted cumulative gain at rank $k$ (DCG@k) is computed as:
\begin{equation}
\mathrm{DCG@}k = \sum_{i=1}^{k} \frac{rel(i)}{\log_2(i+1)},
\label{eq:dcg}
\end{equation}
where $rel(i) \in \{0,1\}$ denotes the relevance of the $i$-th retrieved item. The ideal DCG (IDCG@k) is defined as the maximum obtainable DCG@k:
\begin{equation}
\mathrm{IDCG@}k = \sum_{i=1}^{\min(k,R)} \frac{1}{\log_2(i+1)},
\label{eq:idcg}
\end{equation}
where $R$ is the total number of relevant items for the query. nDCG@k is then obtained by normalizing DCG@k:
\begin{equation}
\mathrm{nDCG@}k = \frac{\mathrm{DCG@}k}{\mathrm{IDCG@}k},
\label{eq:ndcg}
\end{equation}
which ensures that nDCG@k lies in the range $[0,1]$.

\medskip
\textbf{MRR.} Mean Reciprocal Rank (MRR) evaluates the position of the first relevant retrieved item. For a set of $Q$ queries, MRR is computed as:
\begin{equation}
\mathrm{MRR} = \frac{1}{Q} \sum_{i=1}^{Q} \frac{1}{\mathrm{rank}(i)},
\label{eq:mrr}
\end{equation}
where $\mathrm{rank}(i)$ is the position of the first relevant result for the $i$-th query. If no relevant item appears in the retrieved list, we set $\frac{1}{\mathrm{rank}(i)} = 0$. MRR ranges between 0 and 1 and is particularly sensitive to early retrieval performance.

\subsection{Per-Release Retrieval Results}
\label{app:per_release_results}

The main text reports results on \texttt{v202603}; here we provide the corresponding \texttt{v202602} results and a cross-release comparison.
\cref{tab:v202602_results} reports per-task nDCG@10 and Recall@10 for the eleven models evaluated on \texttt{v202602}.

\begin{table}[h]
  \centering
  \small
  \renewcommand{\arraystretch}{1.12}
  \setlength{\tabcolsep}{3.5pt}
  \begin{sc}
  \begin{tabular}{l *{8}{c}}
    \toprule
    \multirow{2}{*}{Model}
      & \multicolumn{2}{c}{Text-to-Code}
      & \multicolumn{2}{c}{Code-to-Text}
      & \multicolumn{2}{c}{Code-to-Code$^\dagger$}
      & \multicolumn{2}{c}{Overall} \\
      \cmidrule(lr){2-3}\cmidrule(lr){4-5}\cmidrule(lr){6-7}\cmidrule(lr){8-9}
      & \multicolumn{1}{c}{nDCG} & \multicolumn{1}{c}{Recall}
      & \multicolumn{1}{c}{nDCG} & \multicolumn{1}{c}{Recall}
      & \multicolumn{1}{c}{nDCG} & \multicolumn{1}{c}{Recall}
      & \multicolumn{1}{c}{nDCG} & \multicolumn{1}{c}{Recall} \\
    \midrule
    C2LLM-7B           & \textbf{0.435} & \textbf{0.765} & \textbf{0.822} & \textbf{0.838} & 0.661 & 0.998 & \textbf{0.639} & \textbf{0.815} \\
    C2LLM-0.5B         & 0.429 & 0.713 & 0.799 & 0.833 & 0.664 & 0.978 & 0.625 & 0.789 \\
    GemEmb-2           & 0.420 & 0.755 & 0.764 & \textbf{0.838} & \textbf{0.709} & \textbf{1.000} & 0.607 & 0.811 \\
    Jina-code-emb-1.5b & 0.405 & 0.713 & 0.767 & 0.827 & 0.686 & 0.976 & 0.600 & 0.786 \\
    F2LLM-4B           & 0.400 & 0.694 & 0.788 & 0.825 & 0.515 & 0.839 & 0.597 & 0.767 \\
    Jina-code-emb-0.5b & 0.397 & 0.679 & 0.742 & 0.808 & 0.699 & 0.980 & 0.585 & 0.761 \\
    Qwen3-Emb-4B       & 0.399 & 0.651 & 0.763 & 0.813 & 0.386 & 0.713 & 0.576 & 0.734 \\
    F2LLM-1.7B         & 0.377 & 0.625 & 0.761 & 0.819 & 0.408 & 0.668 & 0.567 & 0.722 \\
    F2LLM-0.6B         & 0.348 & 0.583 & 0.742 & 0.795 & 0.336 & 0.576 & 0.540 & 0.686 \\
    Qwen3-Emb-8B       & 0.341 & 0.551 & 0.726 & 0.786 & 0.299 & 0.535 & 0.526 & 0.664 \\
    Qwen3-Emb-0.6B     & 0.350 & 0.579 & 0.665 & 0.757 & 0.419 & 0.719 & 0.508 & 0.675 \\
    \bottomrule
  \end{tabular}
  \end{sc}
  \vspace{2pt}
  {\footnotesize $^\dagger$Anchor excluded; Overall is query-count-weighted.}
  \caption{Per-task retrieval results on \textsc{CoREB} \texttt{v202602}. Format matches \cref{tab:per_task_res_table}.}
  \label{tab:v202602_results}
\end{table}

\paragraph{Cross-release comparison.}
For the nine models evaluated on both releases, rankings are largely stable: C2LLM-7B leads on both, and the top-5 share the same members.
Per-task nDCG@10 differences are small (median $|\Delta|$ = 0.02), with no model changing by more than 0.08 on any single task.
Text-to-code scores are slightly higher on \texttt{v202603} for most models, while code-to-text scores are slightly lower; these shifts likely reflect differences in the underlying problem distributions rather than systematic difficulty changes.
The consistency across temporally disjoint releases supports the stability of \textsc{CoREB} as an evaluation instrument.

\subsection{Subtask and Language Analysis}
\label{app:subtask_language_analysis}

\begin{center}
  \includegraphics[width=0.5\linewidth]{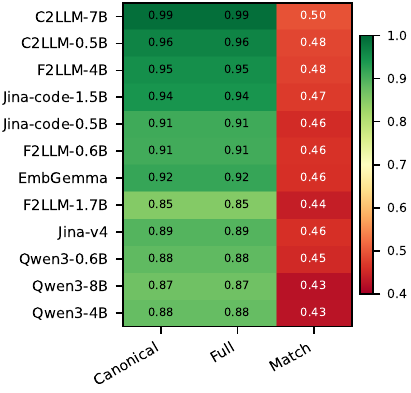}%
  \captionof{figure}{Code-to-text nDCG@10 by subtask type on \texttt{v202603}.}
  \label{fig:c2t_heatmap}
\end{center}

\paragraph{How do models handle short keyword queries?}
The text-to-code Search subtask uses short developer-style queries (19 tokens on average).
On this subtask every model collapses to near-zero nDCG@10, two orders of magnitude below the Canonical subtask (\cref{tab:subtask_breakdown}).
The gap stems from short queries carrying too little context for dense retrieval, not from corpus difficulty (all subtasks share the same 1,744-item corpus).
Going from Canonical to Full (448 tokens) yields at most a marginal delta.
Current models have saturated in the long-query regime while the short-query regime, which most closely mirrors real developer search, remains unsolved.


\paragraph{Does target language affect retrieval quality?}
Text-to-code performance drops sharply when queries specify a target language rather than accepting any language (\cref{fig:lang_gradient}).
Averaged across models and non-search subtasks, language-agnostic queries reach 0.714 nDCG@10, while language-specific constraints reduce this to 0.649 for Python, 0.514 for Java, 0.465 for C++, and 0.383--0.387 for Ruby and Go.
Part of this gap is a confound: ``any'' queries accept solutions in all five languages and therefore have more valid positives per query.
However, a residual bias remains: every model scores lowest on Ruby and Go, regardless of family or scale.
The gradient tracks training-corpus coverage: Python and Java dominate publicly available code, so models embed those languages more faithfully.

\subsection{Per-Subtask and Per-Language Detailed Results}
\label{app:detailed_results}

This section provides fine-grained breakdowns that complement the aggregate numbers in the main text.
\cref{tab:c2c_per_lang} reports code-to-code nDCG@10 stratified by the anchor language, and \cref{tab:corpus_stats} summarizes the corpus composition and token-length statistics.

\begin{table}[h]
  \centering
  \small
  \renewcommand{\arraystretch}{1.12}
  \setlength{\tabcolsep}{5pt}
  \begin{sc}
  \begin{tabular}{l c c c c}
    \toprule
    Model & Python & Java & C++ & Go \\
    \midrule
    GemEmb-2          & 0.681 & 0.674 & 0.789 & 0.673 \\
    Jina-code-0.5b    & 0.690 & 0.493 & 0.788 & 0.684 \\
    Jina-code-1.5b    & 0.688 & 0.499 & 0.790 & 0.650 \\
    C2LLM-7B          & 0.639 & 0.510 & 0.799 & 0.690 \\
    C2LLM-0.5B        & 0.652 & 0.576 & 0.729 & 0.662 \\
    F2LLM-4B          & 0.534 & 0.276 & 0.648 & 0.448 \\
    Qwen3-4B           & 0.498 & 0.174 & 0.504 & 0.186 \\
    Qwen3-0.6B         & 0.488 & 0.147 & 0.445 & 0.244 \\
    F2LLM-1.7B        & 0.485 & 0.137 & 0.512 & 0.192 \\
    F2LLM-0.6B        & 0.482 & 0.105 & 0.357 & 0.113 \\
    Qwen3-8B           & 0.446 & 0.064 & 0.390 & 0.137 \\
    \bottomrule
  \end{tabular}
  \end{sc}
  \caption{code-to-code nDCG@10 by anchor language (after anchor exclusion). Each value is the mean over all queries whose anchor is in the given language.}
  \label{tab:c2c_per_lang}
\end{table}

\begin{table}[h]
  \centering
  \small
  \renewcommand{\arraystretch}{1.2}
  \setlength{\tabcolsep}{5pt}
  \begin{sc}
  \begin{tabular}{l r r r r r r}
    \toprule
    & \multicolumn{2}{c}{\textbf{v202602}} & \multicolumn{2}{c}{\textbf{v202603}} & \multicolumn{2}{c}{\textbf{Aggregated}} \\
    \cmidrule(lr){2-3} \cmidrule(lr){4-5} \cmidrule(lr){6-7}
    \textbf{Corpus} & \#Items & Avg.\ tok & \#Items & Avg.\ tok & \#Items & Avg.\ tok \\
    \midrule
    Code            & 1{,}670 & 371 & 1{,}744 & 401 & 3{,}414 & 386 \\
    \midrule
    Text\textsuperscript{$\ddagger$} &   835 & 687 &   875 & 667 & 1{,}710 & 677 \\
    \quad Original  &   167 & 461 &   175 & 464 &   342 & 463 \\
    \quad Noise     &   668 & 744 &   700 & 718 & 1{,}368 & 731 \\
    \bottomrule
  \end{tabular}
  \end{sc}
  \vspace{2pt}
  \begin{minipage}{\linewidth}
    {\footnotesize\raggedright
    \textsuperscript{$\ddagger$}Text corpora: originals per release (167 / 175) + LLM-generated hard negatives (668 / 700); \emph{Aggregated} columns sum across both releases.}
  \end{minipage}
  \caption{Corpus statistics for \textsc{CoREB} per release and aggregated; token counts from
  \texttt{cl100k\_base} (tiktoken). ``Avg.\ tok'' is the average token length per item, computed over that release's items (and weighted across releases in the aggregated column).}
  \label{tab:corpus_stats}
\end{table}

\subsection{Full Parameter Efficiency Figure}
\label{app:efficiency_full}

\begin{figure}[h]
  \centering
  \includegraphics[width=0.48\linewidth]{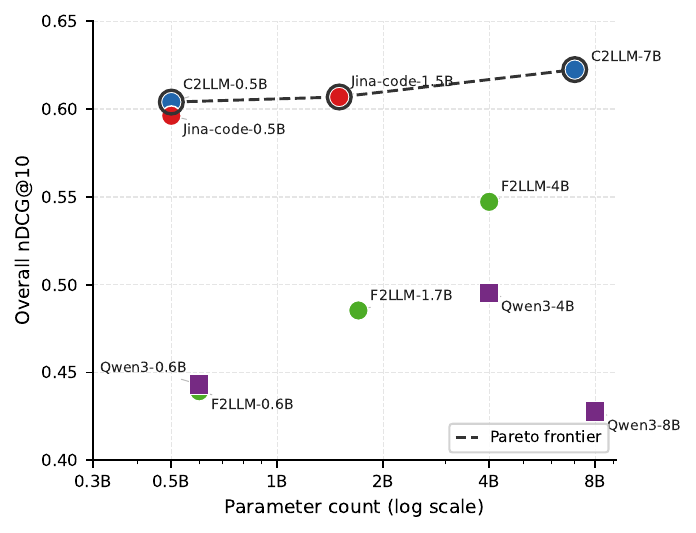}%
  \hfill
  \includegraphics[width=0.48\linewidth]{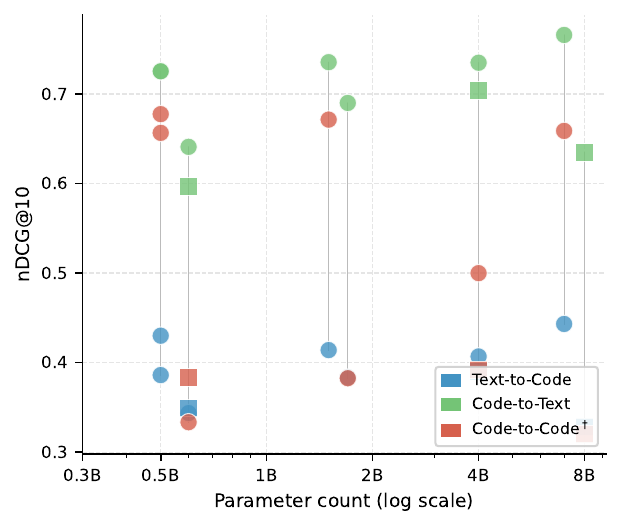}
  \vspace{4pt}
  \includegraphics[width=\linewidth]{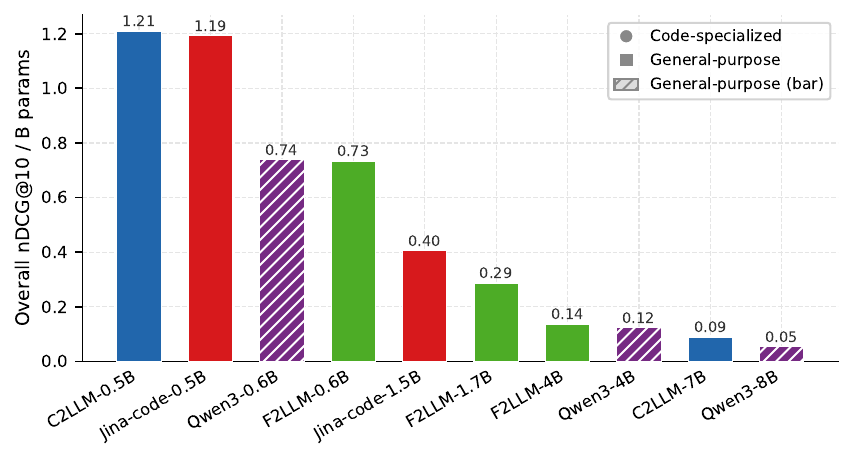}
  \caption{Full model parameter efficiency (expanded version of \cref{fig:efficiency_scatter,fig:efficiency_bar}).
    \textbf{Top-left:} Overall nDCG@10 vs.\ parameter count with all model labels.
    \textbf{Top-right:} Per-task nDCG@10 vs.\ parameter count; vertical lines connect the same model across tasks.
    \textbf{Bottom:} Parameter efficiency (nDCG@10 per billion parameters) for all ten open-weight models, sorted descending. Hatched bars = general-purpose models.}
  \label{fig:efficiency_full}
\end{figure}

\cref{fig:efficiency_full} presents the complete three-panel efficiency analysis.  The main text (\cref{fig:efficiency_scatter,fig:efficiency_bar}) shows only the Pareto frontier scatter plot and the top/bottom efficiency bars; the figure below adds per-task scatter and the full bar ranking.

\subsection{Anchor Exclusion in Code-to-Code Evaluation}
\label{app:anchor_exclusion}

The code-to-code cross-language subtask has a structural property that requires explicit handling in the evaluation.
Each query is a code snippet in a source language; by construction, the anchor code item the query was generated from is also present in the shared retrieval corpus.
Every embedding model assigns near-perfect cosine similarity ($\geq 0.99$) to the anchor, which is byte-for-byte identical to the query, placing it at rank~1 without exception.
However, the anchor is \emph{not} a positive in the qrels (only cross-language translations are labeled relevant), so counting it in the ranked list would assign every model a structurally wasted rank~1, collapsing nDCG@1 and MRR@1 to zero universally and masking genuine differences in cross-language retrieval ability.
To prevent this artifact, \textsc{CoREB}'s code-to-code evaluation removes the anchor from each query's ranked list before computing metrics, so rank~1 in every reported result corresponds to the model's best genuine retrieval candidate.

With anchor exclusion, nDCG@1 becomes a meaningful first-rank precision signal.
Jina-code-0.5b achieves the highest nDCG@1 (0.378) among evaluated models, followed by Jina-code-1.5b (0.356) and GemEmb-2 (0.331); C2LLM-7B reaches 0.327 at rank~1.
By contrast, Qwen3-Emb-8B reaches nDCG@1~$=$~0.155, rarely returning the correct translation first.
The sharp Recall@$k$ curve for code-to-code (\cref{fig:recall_compare}) therefore reflects genuine cross-language retrieval behavior: GemEmb-2 reaches Recall@10~$=$~1.000 on this subtask (with nDCG@10~$=$~0.698), while C2LLM-7B achieves Recall@10~$=$~0.997 (nDCG@10~$=$~0.659) and Qwen3-Emb-8B lags behind (Recall@10~$=$~0.450, nDCG@10~$=$~0.320).
These diverging nDCG and Recall values reveal that the primary bottleneck on this subtask is not retrieval \emph{coverage} but cross-language \emph{ranking precision}: pushing the correct translation to the very top rather than merely within the top-10 window.

The practical implication for benchmark users is straightforward: any evaluation runner for code-to-code tasks that does not exclude the anchor will systematically report inflated Recall (since the anchor trivially satisfies coverage) and deflated precision metrics (since nDCG@1 and MRR@1 are wasted on an irrelevant anchor hit).
\textsc{CoREB} provides the \texttt{c2c\_anchor\_map} (query\,id $\to$ anchor\,code\,id) alongside the task data so that custom evaluation pipelines can reproduce the corrected metrics without re-running the full benchmark.

\subsection{nDCG--Recall Divergence Across Tasks}
\label{app:ndcg_recall}

\begin{center}
  \includegraphics[width=0.5\linewidth]{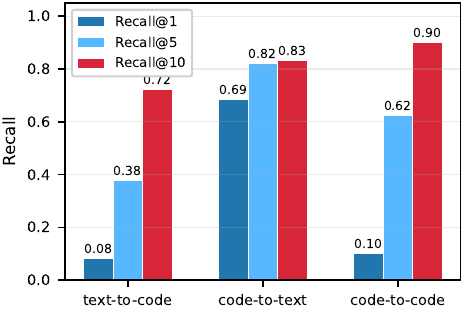}%
  \captionof{figure}{Recall@$k$ at $k\!\in\!\{1,5,10\}$ per task, averaged over all models on \texttt{v202603}. Code-to-code recall rises most sharply from $k\!=\!1$ to $k\!=\!10$, confirming relevant items are retrieved but not top-ranked.}
  \label{fig:recall_compare}
\end{center}

Comparing nDCG@10 and Recall@10 across tasks (\cref{fig:ndcg_compare,fig:recall_compare}) reveals a task-dependent gap between the two metrics.
On code-to-text the metrics track closely, meaning that when a relevant item appears in the top-10 window it also tends to sit near rank one.
On text-to-code and code-to-code the gap is much larger: models do retrieve relevant items within the top ten, but place them at lower ranks, which suppresses nDCG while leaving Recall largely intact.
This pattern points to ranking precision (not coverage) as the primary bottleneck on the harder tasks, and suggests that reranking or listwise training objectives could yield targeted improvements without needing better recall.

\subsection{Extended Related Work}
\label{app:related_work}

\paragraph{LLMs for code.}
Recent progress in both general-purpose and code-specialized LLMs has dramatically improved program synthesis and software engineering assistance, building on early systems such as Codex~\citep{chen2021eval}.
Frontier assistants such as GPT-4.1~\citep{openai2025gpt41} and Claude 3.5~\citep{anthropic2024claude35sonnet} treat coding as a first-class capability, combining strong reasoning, long-context understanding, and agentic workflows.
On the open-source side, StarCoder2~\citep{lozhkov2024starcoder2stackv2} and the Qwen2.5-Coder family~\citep{hui2024qwen25codertechnicalreport} offer 0.5B--32B parameter models trained on multi-language code corpora with strong results across generation, completion, and repair benchmarks.
More broadly, foundation models such as Llama~3~\citep{grattafiori2024llama3herdmodels}, Gemini 3 Pro~\citep{gemini3pro2025}, and Claude 4.5 Sonnet~\citep{claude2025sonnet45} natively support coding alongside multilingual and reasoning capabilities.
These trends highlight a shift from narrow code-only systems toward general LLMs where strong coding competence is integrated into a broader suite of capabilities.

\paragraph{Code embedding benchmarks.}
General-purpose embedding benchmarks such as BEIR~\citep{thakur2021beir} and MTEB~\citep{muennighoff2022mteb} have substantially advanced the evaluation of text and image embeddings but include only a small number of code-related retrieval tasks.
For code specifically, IdBench~\citep{wainakh2021idbench} evaluates embeddings of individual identifier names via human similarity judgments, offering fine-grained insights but focusing on token-level semantics rather than whole-program behavior.
More recent suites, including CoIR~\citep{li2024coir} and CPRet~\citep{deng2025cpret}, extend embedding evaluation to code retrieval across multiple datasets and task types yet remain primarily retrieval-oriented and often reuse heavily studied datasets such as CodeSearchNet~\citep{husain2020codesearchnet} and related variants, raising concerns about overfitting and limited task diversity.
Beyond these issues, these benchmarks exhibit deeper structural problems: universal 1-to-1 qrels that collapse ranking metrics, widespread label noise ($\sim$51\% in CosQA~\citep{gong2026cosqa}), tasks that reduce to string matching or dialogue completion rather than genuine code retrieval, and high contamination risk from datasets that have been public training data for years~\citep{allamanis2019adverse,hernandezlopez2024interdataset}; see \cref{app:coir_qrels,app:coir_cosqa,app:coir_csn,app:coir_ccr,app:coir_cfmt,app:coir_other} for a full dataset-by-dataset analysis.
In contrast, CoREB targets a broader notion of code representation quality by evaluating contamination-limited embeddings across diverse tasks with multi-relevant qrels, explicit hard negatives, and balanced multilingual coverage.

\paragraph{Retrieval models.}
Code retrieval models typically pair a pretrained code encoder with a dense retrieval architecture, mapping natural-language queries and code snippets into a shared embedding space for similarity search.
Early systems build on encoder-only models such as CodeBERT~\citep{feng-etal-2020-codebert} and GraphCodeBERT~\citep{guo2021graphcodebert}, trained with contrastive objectives on NL--code pairs for generic code search.
More recent work introduces large-scale, code-specialized retrievers such as CodeSage~\citep{zhang2024code} and Qodo-Embed~\citep{qodoembed2025}, which leverage web-scale multilingual code corpora and retrieval-aware training pipelines to set state-of-the-art results on benchmarks like CoIR.
However, these models are mostly evaluated on standard code-search suites, and their behavior on the broader set of code- and problem-centric retrieval scenarios we target remains underexplored.

\subsection{Structural Flaws in CoIR}
\label{app:coir_flaws}

CoIR~\citep{li2024coir} aggregates ten code retrieval datasets under a unified evaluation protocol.
While this standardization is valuable, a systematic audit reveals structural issues that affect every constituent dataset and cast doubt on the reliability of CoIR-based model comparisons.
We summarize the universal and dataset-specific flaws below.

\subsubsection{Universal: Trivial 1-to-1 Qrels}
\label{app:coir_qrels}

All ten CoIR datasets assign exactly one relevant document per query, with all relevance scores equal to~1 (no graded relevance, no hard negatives).
The query--document ID suffixes are numerically aligned (e.g., \texttt{qN}$\to$\texttt{dN}), turning retrieval into bipartite matching.
Because the ideal DCG for a query with a single binary-relevant document is always $1/\log_2 2=1.0$, nDCG@$k$ reduces to a function of that document's rank position alone; nDCG and MRR become perfectly correlated, and Recall@$k$ is binary (0~or~1).
The CoIR paper itself acknowledges this limitation: ``each query corresponding to exactly one ground-truth corpus'' fails to capture multi-answer scenarios~\citep{li2024coir}.

\textsc{CoREB} addresses this with multi-relevant qrels: 68\% of text-to-code queries have 2--10 relevant documents (cross-language solutions); code-to-code queries average 2.2 relevant translations; and all three tasks include explicit hard negatives (\texttt{relevance=1}) alongside true positives (\texttt{relevance=2}), totaling 11{,}810 and 12{,}017 graded judgments in \texttt{v202602} and \texttt{v202603} respectively (\cref{tab:graded_qrel_counts}).
These give nDCG and Recall distinct, meaningful roles.

\subsubsection{CosQA: $\sim$51\% Mislabeled Pairs}
\label{app:coir_cosqa}

CosQA pairs Bing web search logs with CodeSearchNet GitHub functions.
This domain mismatch causes systematic mislabeling, independently confirmed by CoSQA+~\citep{gong2026cosqa}, which reports that ``around 51\% of queries are paired with mismatching code.''
Our manual review of 80 pairs found $\sim$37.5\% clear mislabels (code does the opposite or an unrelated thing), $\sim$22.5\% weak/tangential matches, and only $\sim$40\% reasonable matches.
Concrete examples:
\begin{itemize}[nosep,leftmargin=*]
  \item \emph{``python check file is readonly''} is paired with a function testing \texttt{os.access(f, os.R\_OK)} (read permission, the opposite of read-only).
  \item \emph{``how to make seconds to time in python''} is paired with \texttt{time2seconds}, the exact inverse conversion.
  \item \emph{``python making string lower case''} is paired with a CamelCase converter (\texttt{to\_camel}).
\end{itemize}
The test split contains only 500 pairs, of which roughly 200 are reasonable, giving an extremely high-variance evaluation surface.

\subsubsection{CodeSearchNet: Docstring Retrieval, Not Code Search}
\label{app:coir_csn}

CodeSearchNet pairs code with its own docstring.
The original paper acknowledges that documentation ``is often written at the same time and by the same author as the documented code, and hence tends to use the same vocabulary, unlike search queries''~\citep{husain2020codesearchnet}.
ProCQA~\citep{li2024procqa} confirms that the queries are ``either documentation strings or comments rather than natural language questions, limiting its practicality in real scenarios.''
Code2Doc~\citep{karaman2025code2doc} found that aggressive quality filtering retains only 25.6\% of CodeSearchNet-style repository-scraped data, and reports 15--25\% test/train near-duplication.
Because CodeSearchNet has served as public training data since 2019, models evaluated on it face severe contamination risk~\citep{allamanis2019adverse,hernandezlopez2024interdataset}.

\subsubsection{CodeSearchNet-CCR: String Completion}
\label{app:coir_ccr}

CCR randomly splits each CodeSearchNet function at a character position between 40--70\%.
The first half is the query; the second half is the corpus item.
This produces splits mid-docstring sentence, mid-control-flow statement, or mid-variable name.
Both halves share the function name as a metadata \texttt{title} field, providing trivial lexical leakage.
The task measures string continuation memory, not code retrieval ability.

\subsubsection{CodeFeedback-MT: Dialogue Completion}
\label{app:coir_cfmt}

Multi-turn conversations where the query is the full dialogue history (including prior assistant turns with code) and the target is the final assistant response.
The query already contains complete implementations of the same algorithm repeated across turns, enabling near-deduplication matching.
Average query length is 4,558 characters, well beyond the context window of most embedding models.

\subsubsection{Other Datasets}
\label{app:coir_other}

\textbf{CodeFeedback-ST}: Both queries and corpus are LLM-generated, creating artificial stylistic coherence.
Some corpus items contain bare output values (\texttt{[3,4,5]}) or pure prose with zero code, labeled as ``Python.''
\textbf{Synthetic Text2SQL}: Each query includes \texttt{CREATE TABLE} DDL with unique table/column names that appear verbatim in the target SQL; a BM25 baseline scores highly without semantic understanding.
\textbf{StackOverflow QA}: Only $\sim$0.15\% of queries contain code; this is effectively text QA, not code retrieval.
\textbf{APPS}: Python-only, with no quality control; corpus items include personal signatures, opaque variable names, and dead docstrings placed after \texttt{return} statements~\citep{li2023taco,siddiq2024faultinstars}.
\textbf{CodeTransOcean}: Too small for reliable evaluation (180--446 test queries).

\subsubsection{Summary: CoIR vs.\ CoREB}
\label{app:coir_summary}

\begin{table}[h]
\centering
\small
\caption{Structural comparison between CoIR and \textsc{CoREB}.}
\label{tab:coir_vs_coreb}
\begin{tabular}{l p{4.6cm} p{5.2cm}}
\toprule
\textbf{Dimension} & \textbf{CoIR} & \textbf{CoREB} \\
\midrule
Qrel structure & 1-to-1, score $=$ 1 only & Multi-relevant + graded hard negatives (relevance $=$ 1) \\
Label quality & $\sim$51\% mislabeled (CosQA) & Programmatically verified \\
Languages & Python $+$ SQL only (meaningful tasks) & 5 languages, balanced across all tasks \\
Contamination risk & High (public since 2019) & Low (fresh LLM-generated, Feb 2025) \\
Hard negatives & None across all 10 datasets & 668/700 noise corpus items per release (1{,}368 aggregated) + relevance-1 qrels \\
\bottomrule
\end{tabular}
\end{table}

\subsection{Limitations}
\label{app:limitations}

\textsc{CoREB} inherits several scope constraints.
First, the corpus is drawn exclusively from competitive programming via LiveCodeBench, so the benchmark may not fully represent retrieval over enterprise or library code.
Second, only five programming languages are covered; extending to lower-resource languages (e.g., Rust, Kotlin) could change the difficulty landscape.
Third, all queries are LLM-generated from problem statements rather than collected from real developer search logs, which limits ecological validity for the short-query setting.
Finally, the current evaluation is offline and single-turn; interactive or multi-turn code search scenarios are not addressed.


\end{document}